\documentclass[twocolumn,aps,pra]{revtex4-1}
\usepackage{amsfonts}
\usepackage{comment}
\usepackage{amsmath}
\usepackage{amssymb}
\usepackage{graphicx}
\usepackage{color}
\providecommand{\U}[1]{\protect\rule{.1in}{.1in}}
\providecommand{\U}[1]{\protect\rule{.1in}{.1in}}
\providecommand{\U}[1]{\protect\rule{.1in}{.1in}}
\providecommand{\U}[1]{\protect\rule{.1in}{.1in}}
\providecommand{\U}[1]{\protect\rule{.1in}{.1in}}
\providecommand{\U}[1]{\protect\rule{.1in}{.1in}}
\providecommand{\U}[1]{\protect\rule{.1in}{.1in}}
\providecommand{\U}[1]{\protect\rule{.1in}{.1in}}
\providecommand{\U}[1]{\protect\rule{.1in}{.1in}}
\providecommand{\U}[1]{\protect\rule{.1in}{.1in}}
\providecommand{\U}[1]{\protect\rule{.1in}{.1in}}

\providecommand{\U}[1]{\protect\rule{.1in}{.1in}}
\providecommand{\U}[1]{\protect\rule{.1in}{.1in}}
\providecommand{\U}[1]{\protect\rule{.1in}{.1in}}
\providecommand{\U}[1]{\protect\rule{.1in}{.1in}}
\providecommand{\U}[1]{\protect\rule{.1in}{.1in}}
\providecommand{\U}[1]{\protect\rule{.1in}{.1in}}
\providecommand{\U}[1]{\protect\rule{.1in}{.1in}}
\providecommand{\U}[1]{\protect\rule{.1in}{.1in}}

\def\calh{\mathcal{H}}

\def\R{\mathbb{R}}

\def\E{{\rm I\kern-.1567em E}}
\def\A{{\rm I\kern-.1567em A}}
\def\P{{\rm I\kern-.1567em P}}
\def\V{{\rm I\kern-.1567em V}}

\def\bq{\begin{equation}}
\def\eq{\end{equation}}
\def\bqy{\begin{eqnarray}}
\def\eqy{\end{eqnarray}}

\def\de{\delta}

\def\ga{\gamma}

\def\ka{\kappa}
\def\la{\lambda}

\def\na{\nabla}

\def\Om{\ze}

\def\si{\sigma}

\def\ze{\zeta}

\def\bfb{\mathbf{b}}
\def\bfB{\mathbf{B}}

\def\bfF{\mathbf{F}}
\def\bfa{\mathbf{a}}

\def\bfL{\mathbf{L}}
\def\bfP{\mathbf{P}}
\def\bfQ{\mathbf{Q}}

\def\bfV{\mathbf{V}}
\def\bfx{\mathbf{x}}
\def\bfv{\mathbf{v}}

\def\bfq{\mathbf{q}}

\def\bfxi{\boldsymbol{\xi}}

\def\bfet{\boldsymbol{\eta}}

\def\bfPi{\boldsymbol{\Pi}}
\def\bfpi{\boldsymbol{\pi}}
\def\bfla{\boldsymbol{\la}}
\def\bfOm{\boldsymbol{\Om}}

\def\mfA{\mathfrak{A}}
\def\mfB{\mathfrak{B}}
\def\mfJ{\mathfrak{J}}

\def\p{\partial}

\def\na{\nabla}

\def\inta{\int\! d^{3}a \,}
\def\intb{\int\! d^{3}b \,}
\def\intx{\int\! d^{3}x \,}

\begin{document}


\title{Hamiltonian Magnetohydrodynamics:\\ Lagrangian, Eulerian,  and Dynamically
Accessible Stability - Theory}
\author{T. Andreussi}
\affiliation{Alta S.p.A., Pisa, 56121, Italy}
\author{P. J. Morrison}
\affiliation{Institute for Fusion Studies and Department of Physics, The University of
Texas at Austin, Austin, TX 78712-1060, USA}
\author{F. Pegoraro}
\affiliation{Dipartimento di Fisica E.~Fermi, Pisa, 56127, Italy}
\date{\today}

\begin{abstract}
Stability conditions of magnetized plasma flows are obtained by exploiting the Hamiltonian structure of the magnetohydrodynamics (MHD) equations and, in particular, by using three kinds of energy principles. First, the Lagrangian variable  energy principle is described  and sufficient stability conditions are presented. Next, plasma flows are described in terms of Eulerian variables and the noncanonical Hamiltonian formulation of MHD is exploited. For symmetric equilibria, the energy-Casimir principle is expanded to second order and sufficient conditions for stability to symmetric perturbation are obtained. Then, dynamically accessible variations, i.e. variations that explicitly preserve  invariants of the system, are introduced and the respective energy principle is considered. General criteria for stability are obtained, along with  comparisons between the three different approaches.

\end{abstract}
\maketitle


\section{Introduction}
\label{sec:intro}

In this paper, a companion to Ref.~\cite{amp1} and its predecessor Ref.~\cite{amp0}, we explore further ramifications of the Hamiltonian nature of ideal magnetohydrodynamics (MHD).  Whereas in  Refs.~\cite{amp1,amp0}  the subject matter concerned the construction and origin of variational principles for equilibria, here we present the comprehensive approach to stability of  magnetohydrodynamics (MHD) equilibria that is a direct consequence of the Hamiltonian nature of this system.  The presentation organizes scattered  approaches into the cohesive Hamiltonian framework, which will be seen to be useful for obtaining, interpreting,  and comparing stability results

Ultimately, the  stability results we consider, which are a consequence of  the Hamiltonian form,  have their origin in  two energy theorems of mechanics:  Lagrange's theorem and Dirichlet's theorem (see Ref.~\cite{morrison98} for review).  The former is the root of the necessary and sufficient hydromagnetic energy principle  of Refs.~\cite{Bernstein1958b,lust-shluter,hain} for static equilibria, while the latter is the root of various Eulerian sufficient conditions for stability (see e.g, Ref.~\cite{morrison-eliezer}).  Since MHD, being a set of partial differential equations, is an infinite-dimensional Hamiltonian system, there are technical aspects not present in the theorems of mechanics.  For example, MHD can be expressed in terms of the Lagrangian or  Eulerian  variable descriptions, each of which enforces constraints in particular but nonequivalent ways.   A main goal of this paper is to explore the consequences for stability for  such different ways of enforcing constraints.  The results of \cite{amp1,amp0} will be used in Secs.~\ref{sec:lag}, \ref{sec:ec}, and \ref{sec:da} to construct three kinds of energy principles for stability of both static and stationary MHD equilibria.  In these sections results of a more general nature are obtained, while  more specific examples and comparisons will be made in a companion paper \cite{amp2b}.

More specifically, in Sec.~\ref{sec:lag}, energy stability in the purely Lagrangian variable framework, as considered in Ref.~\cite{Frieman1960},  will be treated. This form extends the classical  hydromagnetic energy principle of \cite{Bernstein1958b}, obtained for static  configurations, to stationary flows.   En route to our results we briefly do the following: (i) review the Hamiltonian description in terms of Lagrangian variables and  describe a time-dependent relabeling transformation, which to our knowledge has not heretofore been given, a transformation that will be needed for later development, (ii) review the map from Lagrangian to Eulerian variables, so as to understand how  the time dependence of  stationary equilibria in the Lagrangian picture relates to  time-independent Eulerian equilibria and how such time dependence can be removed,  and (iii) expand Lagrangian particle trajectories to obtain energy expressions for perturbations of general equilibria and use these expressions for obtaining sufficient conditions for stability of equilibria with  stationary flow.


In Sec.~\ref{sec:ec} the second kind of energy principle is described, one that has purely
Eulerian form in terms of the usual MHD variables. This form has been called
the energy-Casimir method (see e.g.\ \cite{hazeltine, holm,morrison-eliezer,morrison98}),
although the method predates the name and, in fact, it appeared in the early
plasma literature in several contexts, the earliest of which appears to be
\cite{kruskal-oberman}. This energy principle gives sufficient conditions for
stability by expanding a functional $F$ composed of the sum of the Eulerian
energy plus Casimir invariants, an example being the cross helicity $\intx \mathbf{v}\cdot\mathbf{B}$,   to second order. If this second variation is sign definite, then $F$ serves as a Lyapunov functional for stability. With
this energy principle we can assess the stability of equilibria within the
assumed symmetry class.  In this context,  very general and new stability conditions are obtained.

Next, in Sec.~\ref{sec:da} the third kind of energy principle, one that uses
dynamically accessible variations,  terminology for a concept introduced in
\cite{morrison-pfirsch} for a general class of variations generated from the
noncanonical Poisson bracket and consequently explicitly preserves invariants
of the system,  is described. (See \cite{morrison98} for review.) Dynamically accessible
variations do not rely on any symmetry and thus give general criteria for
stability. Therefore, they provide information about the generality of our
second class of energy principles.


Finally, in  Sec.~\ref{sec:conclu} we summarize and conclude.  Here we discuss our results, state implications about
nonlinear stability, and  make comparisons between the various kinds of stability.   Comparisons are made on a general level, which is  somewhat complicated,  but this will pave the way for the companion paper of Ref.~\cite{amp2b}, where a collection of more  specific examples  will be treated and explicit comparisons made.




\section{Lagrangian stability}
\label{sec:lag}

\label{sec:genHamRelab}

\subsection{General Hamiltonian form, relabeling, and conservation laws}
\label{ssec:ghrcs}

Consider a general Hamiltonian field description in terms of a configuration
field, $\bfq\in\R^3$, which subsequently will be the Lagrangian variable that determines the position of a fluid element.  The usual three-dimensional spatial domain is assumed, but the treatment of this subsection applies to any number of dimensions.  Suppose $\bfq$ has a canonical conjugate $\bfpi$ and both are labeled by a continuum variable $\bfa\in\R^3$, i.e., the dynamical variables of the Hamiltonian description  are the pair $(\mathbf{q}(\mathbf{a},t),\boldsymbol{\pi}(\mathbf{a},t))$.  It is common to assume  that the fluid element described by  $\bfq$ is labeled by its initial condition, $\bfq(0,t)=\bfa$, but as will soon become evident this is not necessary.  The phase space of this setup is sometimes denoted by $T^*Q$, where $Q$ is the set of smooth invertible mappings of the spatial domain, indicated here by $\bfq$,  and $T^*Q$ denotes the space (cotangent bundle) with coordinates $\bfq$ together with their conjugate momenta $\bfpi$.  Because the infinite-dimensional geometry implied by $T^*Q$ is backed by meager mathematical rigor,  the language of  Lagrange \cite{lagrange} and  Newcomb \cite{newcomb} will be used here, except because general curvilinear coordinates may be  employed indices will be placed  as in \cite{morrison98,padhye1,padhye2}  indicating  their tensorial character, viz.,  $\mathbf{q}\rightarrow q^i$ and $\boldsymbol{\pi}\rightarrow \pi_i$, where $i=1,2,3$.

%

In terms of the canonical coordinates, $q^{i}$, and momentum densities, $\pi_{i}$, the  dynamics
can be written as%
\begin{equation}
\dot{\pi_{i}}=\left\{  \pi_{i},H\right\} =-\frac{\de H}{\de q^i} \quad {\rm and} \quad
\dot{q}^i=\left\{q^i,H\right\} =\frac{\de H}{\de \pi_i} \,  ,
\label{eq:Hamilton}%
\end{equation}
where `$\ \dot{\ }\ $'  means derivative with respect to $t$ at fixed  label $\bfa$ and the Poisson bracket $\left\{  \cdot,\cdot\right\}  $\ is canonical and  given by
\begin{equation}
\left\{ F,G\right\}  =\inta  \left(  \frac{\delta F}{\delta q^i}%
\frac{\delta G}{\delta \pi_i}-\frac{\delta G}{\delta q^i}%
\frac{\delta F}{\delta \pi_i}\right)\,.
\label{cbkt}%
\end{equation}
In (\ref{cbkt}) $F$ and $G$ are functionals,  repeated indices are to be summed,  and  ${\delta F}/{\delta q^i}$ denotes the  functional derivative of $F$ with respect to $q^i$ (see e.g.\ \cite{morrison98}). Given a Hamiltonian functional of the form
\bq
H[\mathbf{q},\boldsymbol{\pi}]=\inta
\calh(\mathbf{q},\boldsymbol{\pi},\nabla_a \mathbf{q},\nabla_a \boldsymbol{\pi},\dots ,\boldsymbol{a},t)\,,
\label{Hgen}
\eq
where $\nabla_a :=\p/\p \bfa$,  Eqs.~(\ref{eq:Hamilton}) and (\ref{cbkt}) imply a set of partial differential equations.

Consider general transformations of such Hamiltonian systems under an arbitrary time-dependent, invertible relabeling
\bq
\bfa=\mathfrak{A}(\bfb,t) \quad \longleftrightarrow \quad \bfb=\mathfrak{B}(\bfa,t)\,;
\label{relabel}
\eq
i.e.\  $\mfA=\mfB^{-1}$.  It should be emphasized that the transformation of $\mfB$ and its inverse is not connected at this point to the dynamics in any way, nor is it related to symmetry as in \cite{newcomb67, padhye1,padhye2}.    This kind of label change does not usually appear  in traditional finite-dimensional Hamiltonian theory, since it would  amount to a time-dependent change of the  label  $i$ of, e.g.,  a  canonical coordinate $q^i(t)$.  
However, this relabeling transformation is in fact a time-dependent canonical transformation induced by
$\bfQ(\bfb,t)=\bfq(\mfA(\bfb,t),t)$, the transformation to the new coordinate.  To understand how the associated momentum transforms, the following type-2 time-dependent generating functional is used:
\bq
F_2[\mathbf{q},\boldsymbol{\Pi},t]=\!\inta \!\! \intb \,  \bfq(\bfa,t) \cdot  \bfPi(\bfb,t)\,  \de(\bfa-\mfA(\bfb,t))\,,
\label{gengen}
\eq
where $\de$ is the Dirac delta function. (In tensorial form $ \bfq \cdot  \bfPi=q^i\,  \Pi_i$.)  The  direct transformation  from the Hamiltonian theory in terms of  $(\mathbf{q}(\mathbf{a},t),\boldsymbol{\pi}(\mathbf{a},t))$ to that in terms of
$(\mathbf{Q}(\mathbf{b},t),\boldsymbol{\Pi}(\mathbf{b},t))$ is  given by
\bq
\bfpi=\frac{\de F_2}{\de \bfq}\,,\qquad \bfQ=\frac{\de F_2}{\de \bfPi}\,,
\qquad {\rm and}\quad \tilde H =H + \frac{\p F_2}{\p t}\,.
\label{geneqs}
\eq
From (\ref{gengen})  it  follows that
\bq
\bfpi(\bfa,t)=\frac{\bfPi(\bfb,t)}{\mfJ}\,,\qquad\bfQ(\bfb,t)=\bfq(\bfa,t)\,,
\label{genlab}
\eq
where $ \de(\bfa-\mfA)= \de(\bfb-\mfA^{-1})/\mfJ$ has been used, and
\bq
\frac{\p F_2}{\p t}= - \intb\,  \bfPi\cdot (\bfV \cdot \nabla_b \bfQ)
= - \intb\,  \Pi_i  V^j \frac{\p Q^i}{\p  b^j} \,
\label{genlabt}
\eq
where the determinant  $\mfJ:= \left|\p a^i/\p b^j\right|$, which means  $d^3a=\mfJ\,   d^3b $,  and
 \bq
\bfV(\bfb,t):= \dot\mfB\circ\mfB^{-1}=\dot\mfB(\mfA(\bfb,t),t)\,.
\label{VR}
\eq
Equation (\ref{genlabt}) follows from
$
 ({\p \mfA^i}/{\p b^{k}})({\p \mfB^{k}}/{\p a^j})=\de^i_{j}
$
and  
\bq
\frac{d }{dt}\mfB^k(\mfA(\bfb,t),t)= \dot{\mfB}^k + \frac{\p \mfB^k}{\p a^j}\,  {\p_t \mfA^j} = \frac{d b^k}{dt}=0 \,. 
\label{id2}
\eq
Recall `$\ \dot{\ }\ $' always means time differentiation at fixed $\bfa$, while   $\p_t$ will  mean at fixed $\bfb$.
The formulas of (\ref{genlab}) and  (\ref{genlabt})  are valid with substitution of either $\bfa$ or $\bfb$,  using  (\ref{relabel}).

Because the transformation is generated by $F_2$,  it  is a canonical transformation, i.e., the Poisson bracket becomes
\bq
\left\{F,G\right\}
= \intb \left(  \frac{\delta F}{\delta {Q^i}}
\frac{\delta G}{\delta{\Pi_i}}-\frac{\delta G}{\delta{Q^i}}
 \frac{\delta F}{\delta {\Pi_i}}\right)\,,
 \label{QPibkt}
\eq
the Hamiltonian in the new variables becomes
\bq
\tilde H[\bfQ,\bfPi]= H - \intb\,  \bfPi\cdot (\bfV \cdot \nabla_b \bfQ)\,,
\label{newHam}
\eq
The second term of (\ref{newHam}), the one that comes from $\p F_2/\p t$,  will be referred to as the {\it fictitious} term.  The transformed equations of motion are given by
\begin{equation}
\p_t{\Pi_{i}}=\left\{ \Pi_{i},\tilde{H}\right\} \qquad {\rm and}\qquad
\p_t{Q}^i=\left\{Q^i,\tilde{H}\right\} \,.
\label{eq:QPiHamilton}%
\end{equation}

The relabeling transformation of (\ref{relabel}) can also be interpreted  as transformation to a moving noninertial frame of reference.  With this interpretation, $\bfq$ describes motion relative to states  in the  inertial frame with coordinates $\bfa$, and the relabeling transformation amounts to transformation to a noninertial frame with $\bfQ$ describing motion relative to the frame with coordinates $\bfb$.    This explains why relabeling gives rise to the presence of the fictitious  (noninertial) term in the Hamiltonian.  It is important to reiterate that $\bfq$, $\bfQ$ $\mfB$, and $\mfA$ are all invertible maps (parameterized by time) defined on the  same configuration space.

In the case where $H$ is time-independent, energy is conserved, i.e., $\dot H=0$. If $H$ has no explicit  dependence on $\bfq$, but depends on  $\nabla_a\bfq$ and possibly higher derivatives, the momentum $\bfP:=\inta  \bfpi$
is conserved.  This momentum expression inserted into the Poisson bracket  generates an operator for space translations.  Conservation of $\bfP$ follows from
\bq
\dot\bfP=\{\bfP, H\}= -\inta \frac{\de H}{\de \bfq}=\inta \nabla_a(\dots) =0\,,
\eq
where the last equality is true for any functional  that depends on first and higher order derivatives of $\bfq$.
Similarly, for isotropic Hamiltonians  the angular momentum, $\bfL=\inta \bfq\times\bfpi$,
can be shown to be  conserved, which is the case for the MHD Hamiltonian.  When $\bfL$ is inserted into  the Poisson bracket an expression for  the operator that generates rotations is obtained.    The Hamiltonian with these invariants and another, the position of the center of mass that generates Galilean boosts, together with the Poisson bracket,  are a realization of the algebra of the ten parameter Galilean group (see \cite{morrison82}).

 {In terms of the relabeled coordinates the same transpires.  Although a time-independent $H$ may obtain explicit time dependence when written in terms of $(\bfQ,\bfPi)$ and likewise $\tilde H$,  constants of motion remain constants of motion.  For example, the momentum $\bfP$ written in terms of the relabeled coordinates becomes $\bfP= \intb \bfPi$, and because $\de /\de \bfQ$ of the fictitious term of (\ref{newHam})  is  still of the form $\nabla_b(...)$,  it follows that $\{\bfP, \tilde H\}=0$.   Similarly, the  angular momentum remains conserved.   Thus, upon relabeling the set of invariants, with the new Poisson bracket, remain a realization of the algebra of the Galelian group.  This is to be expected since the  Eulerian description does not see the labels and the Eulerian constants of motion with the noncanonical Poisson bracket are a realization of the Galelian group.}

When systems have symmetry one can transform  into  a new frame of reference.  When doing so,  the Hamiltonian generally changes because the transformation is a time-dependent canonical transformation.  For example, using momentum conservation the old Hamiltonian $H$ becomes $H_{\la}= H +\bfla\cdot\bfP$
where the parameter $\bfla$ determines the speed of the  translating frame.  Extremals  of $H_{\la}$  are equilibria in the translating frame and, thus,  correspond to uniformly translating states in the original frame.  Similarly,  equilibria in rotating  frames are extremals of $H_{\Om}=H +\bfOm\cdot \bfL$, where $\bfOm$ determines the magnitude and direction of the rotation.   Such Hamiltonian shifts can be used to obtain stability results for a larger class of states.


\subsection{MHD and the Lagrange-Euler map}
\label{ssec:mhdlemap}

The Hamiltonian for MHD lies in the class of  so-called `natural' Hamiltonians of the form
\begin{equation}
H[\mathbf{q},\boldsymbol{\pi}]=\inta\left[  \frac{|\boldsymbol{\pi}|^2}{2\rho
_{0}}+\rho_{0}\mathcal{W}\left(\mathbf{q},\nabla_a \mathbf{q},\dots ,\boldsymbol{a}\right) \right]\,,
 \label{eq:H_LagrW}
\end{equation}
where $\mathcal{W}$ is some potential energy density and $\rho_0=\rho_0(\bfa)$ is a given function that denotes the mass density of a Lagrangian fluid element.  In a general coordinate system $|\boldsymbol{\pi}|^2=  g^{ij}(\bfq) \, {\pi}_i\,  {\pi}_j=: {\pi}^i \pi_i$, while for Cartesian coordinates the metric is $\eta_{ij}=\eta^{ij}=\de_{ij}$, the usual Kronecker symbol.

The specific form of the Hamiltonian for MHD must satisfy the Eulerian closure principle described in \cite{morrisonaip};  that is,  it must be expressible in terms of the Eulerian variables of the theory. Using the notation of  Ref.~\cite{amp1}   the set  of  Eulerian variables for MHD is denoted by $Z:=\left(  \rho,\mathbf{v},s,\mathbf{B}\right)$, or alternatively 
$\mathcal{Z}:=\left(\rho,\mathbf{M}:=\rho\mathbf{v},\si:=\rho s,\mathbf{B} \right)$,   with the map from the Lagrangian variables $\left( \mathbf{q},\boldsymbol{\pi}\right) $ to Eulerian variables $Z$ given by
\begin{align}
\rho\left(  \mathbf{x},t\right)   &  =\left.  \frac{\rho_{0}\left(  \mathbf{a}\right)  }{J\left(  \mathbf{a}%
,t\right)  }\right\vert_{\mathbf{a}=\mathbf{q}^{-1}\left(  \mathbf{x},t\right)  },\\
v_i\left(  \mathbf{x},t\right)   &  =\left.  \frac{\pi_{i}\left(  \mathbf{a},t\right)  }{\rho_{0}\left(
\mathbf{a}\right)  }\right\vert _{\mathbf{a}=\mathbf{q}^{-1}\left(  \mathbf{x},t\right)  },\\
s\left(  \mathbf{x},t\right)   &  =\left.  s_{0}(\bfa)\right\vert _{\mathbf{a}%
=\mathbf{q}^{-1}\left(  \mathbf{x},t\right)  },\\
B^{i}\left(  \mathbf{x},t\right)   &  =\left.  \frac{\partial q_{i}\left(  \mathbf{a},t\right)  }{\partial a_{j}%
}\frac{B_{0j}\left(  \mathbf{a}\right)  }{J\left(  \mathbf{a},t\right)
}\right\vert_{\mathbf{a}=\mathbf{q}^{-1}\left(
\mathbf{x},t\right)  },
\label{ELmap}
\end{align}
where on the right-hand side $\rho_{0},\ s_{0}$ and $B_{0}^i$  are respectively the plasma density,
the entropy per unit mass and the  $i$th-component of the magnetic field, and
 the subscript zero  indicates that these functions are attributes of the  Lagrangian fluid elements  and thus depend on the label $\mathbf{a}$.   The left-hand side gives the set of usual  Eulerian variables with $\rho$  the plasma density,
$\mathbf{v}$ the flow velocity, $\mathbf{B}$ the magnetic field,  and $s$ the entropy per unit mass that are functions  of the  Eulerian observation position  $\mathbf{x}$.    The Lagrange-Euler map is effected using
$\mathbf{a}=\mathbf{q}^{-1}\left(  \mathbf{x},t\right)$ (the function $\mathbf{q}^{-1}\left(  \mathbf{x},t\right)$, the
inverse of $\mathbf{q}(\mathbf{a},t)$,  indicates the label of the particle
that, at time $t$, is located at the observation point $\bfx$).   Here the determinant $J:=\left| \p q^i/\p a^j\right|$ should
not be confused with $\mfJ= \left|\p a^i/\p b^j\right|$ introduced in Sec~\ref{ssec:ghrcs}.   See Refs.~\cite{newcomb,morrison98,padhye1,padhye2,morrison05} for more details.

The Lagrange-Euler map can also be used to express the variables  $Z$ in terms of the relabeled canonical coordinates $(\bfQ,\bfPi)$.  For example, the entropy per unit mass that will be observed at point $\bfx$ at time $t$ will be that attached to the fluid element there then; hence,  it is gotten by  solving
$\bfx=\bfQ(\bfb,t)$ for $\bfb$,  giving $\bfb=\bfQ^{-1}(\bfx,t)$  and
\bqy
s(\bfx,t)&=&\left.s_0(\bfa,t)\right|_{\bfa=\bfq^{-1}(\bfx,t)}
= \left.s_0(\mfA(\bfb,t),t)\right|_{\bfb=\bfQ^{-1}(\bfx,t)} \nonumber\\
&=:&\left.\tilde{s}_0(\bfb,t)\right|_{\bfb=\bfQ^{-1}(\bfx,t)}\,.
\label{tildes}
\eqy

Similarly, the Eulerian velocity $\bfv(\bfx,t)$, when represented in terms of the new variables, is still the velocity of the fluid element that is at the observation point $\bfx$ at time $t$, but now given in terms of  $\p_t \bfQ(\bfb,t)$,
 \bqy
\bfv(\bfx,t)&=&\left.\dot \bfq(\bfa,t)\right|_{\bfa=\bfq^{-1}(\bfx,t)}
\nonumber\\
&=&\Big(\p_t \bfQ(\bfb,t) \nonumber\\
&{\ }& \qquad + \dot \mfB(\mfA(\bfb,t),t)\cdot \nabla_b \bfQ(\bfb,t)\Big)
\Big|_{\bfb=\bfQ^{-1}(\bfx,t)} \nonumber\\
&=&\left.\left( \p_t\bfQ + \bfV \cdot \nabla_b \bfQ\right)\right|_{\bfb=\bfQ^{-1}(\bfx,t)}\,.
\nonumber
\eqy
Here, the second term comes from label advection.


Mass conservation implies $\rho_0d^3a= \tilde \rho_0 d^3b$,  and thus  $\tilde \rho_0 =\mfJ\,  \rho_0$.  Evidently,  $\tilde J:=|\p Q^i/\p b^j| = J \mfJ$  and $\tilde\rho_0/\tilde J= \rho_0/ J$.  In terms of the relabeled coordinate,  the Eulerian density is
\bq
\rho(\bfx,t) =\left.\frac{\rho_0}{J}\right|_{\bfa=\bfq^{-1}(\bfx,t)}
= \left.\frac{\tilde\rho_0}{\tilde J}\right|_{\bfb=\bfQ^{-1}(\bfx,t)}\,,
\label{tildeden}
\eq
where composition of arguments is as in (\ref{tildes}).

Finally, the magnetic field is similarly expressed as
\bq
B^{i}\left(  \mathbf{x},t\right)    =\left.  \frac{\partial Q^{i}}{\partial
b^{j}}\frac{\tilde{B}_{0}^j}{\tilde{J}}\right\vert _{\mathbf{b}=\mathbf{Q}^{-1}\left(
\mathbf{x},t\right)  }\,.
\label{tildeB}
\eq

The usual equations of motion for $Z$ follow from either of the expressions in terms of $\bfq$ or $\bfQ$ (see \cite{amp1}).  The notation  $\p /\p t$ will be used to  denote differentiation of Eulerian quantities at fixed $\bfx$.

The Hamiltonian for MHD is
\bqy
H[\mathbf{q},\boldsymbol{\pi}]&=&\inta
\Big[
\frac{ \pi_{i} \pi^{i}}{2\rho_{0}}
+ \rho_{0}U\left(  s_{0},\rho_{0}/J\right)\nonumber\\
 &{\ }& \hspace{.75 in} + %
\frac{\partial q_{i}}{\partial a^{k}}  \frac{\partial q^{i}}{\partial a^{\ell}}
\frac{B_{0}^kB_{0}^{\ell}}{8\pi J}
\Big]\,.
\label{eq:H_Lagr}%
\eqy
The function $U$ of (\ref{eq:H_Lagr}) is  the internal energy per unit mass of the plasma. As it is written, it can be expressed as a function of $\rho$ and $s$, i.e. $U=U\left(s,\rho\right)$; this is necessary for this Hamiltonian to satisfy the Eulerian closure principle \cite{morrisonaip}, which in this case means that upon substitution of (\ref{ELmap}), (\ref{eq:H_Lagr}) becomes
\bq
H=\intx
\left[
\frac{\rho }{2}|\bfv|^2
+ \rho \, U(s,\rho) + \frac{ |\bfB|^2}{8\pi}
\right]\,,
\label{eq:H_Euler}%
\eq
an expression entirely in terms of the variables $Z$.  With  the usual
thermodynamic relations,   the pressure is given  $p=\rho^{2}{\partial
U}/{\partial\rho}$ and the temperature by  $T={\partial U}/{\partial s}$.   For  polytropic equations of state, $p=\kappa(s)\rho^{\gamma}$,  $U=\kappa(s)\rho^{\gamma -1}/(\ga -1)$ and with this choice the internal energy integrand of (\ref{eq:H_Euler}) is $\rho  U = p/(\ga -1)$.  Isothermal processes ($\ga=1$) have $U=\ka \ln(\rho)$.

The MHD model can be generalized by altering the Hamiltonian in many physically meaningful ways:  for example,  an anisotropic pressure tensor can be treated as in \cite{morrison82,thermo13} by assuming $U$ depends on $B=|\bfB|$ with
\bq
p_{||}=\rho^2 \frac{\p U}{\p \rho} \quad {\rm and}\quad
p_{\perp}= \rho^2 \frac{\p U}{\p \rho}  + \rho B \frac{\p U}{\p B} \,,
\eq
which gives the CGL equations \cite{cgl},  and the effects of a gravitational force can be modeled by adding to the integrand of (\ref{eq:H_Lagr}) a term $\rho_{0}\varphi$, where
$\varphi$  is an  external potential.

 Now consider explicitly the effect of the relabeling transformation of (\ref{relabel}) on  the MHD Hamiltonian, which we write out in tensorial form
 \bqy
\tilde H[\bfQ,\bfPi]
&=&\intb \Big[
\frac{ \Pi_i \, \Pi^i}{2\tilde\rho_0}
-  \Pi_i  V^j \frac{\p Q^i}{\p b^j} \nonumber\\
&+& \tilde\rho_0\,U\left(\tilde s_0,\tilde\rho_0/\tilde J\right)
 +
 \frac{\partial Q_{i}}{\partial b^{k}}  \frac{\partial Q^{i}}{\partial b^{\ell}}
\frac{\tilde{B}_{0}^k \tilde{B}_{0}^{\ell}}{8\pi \tilde{J}}
\Big]\,,
\nonumber
\\
&=& K + H_f + W\,,
\label{tildeH_MHD}
\eqy
where $K$ is the kinetic energy, $H_f$ is the fictitious term of (\ref{genlabt}), and $W$ represents the sum of the internal and magnetic field energies.
The Hamiltonians of (\ref{eq:H_Lagr}) and (\ref{tildeH_MHD}) and the brackets of (\ref{cbkt}) and (\ref{QPibkt})  are the starting point for the equilibrium and stability analysis of  the next section.

\subsection{Lagrangian description of  equilibrium and stability}
\label{ssec:equil-stab}

Since the theory as thus far described is canonical, Lagrangian variable equilibria are given by ${\delta
H}/{\delta}\mathbf{q}=0$ and ${\delta H}/{\delta}\boldsymbol{\pi}=0$. The
second of these conditions clearly implies $\boldsymbol{\pi}=0$, which means Lagrangian equilibria correspond to  static configurations in which fluid particles do not move.  Thus Eulerian stationary equilibria, i.e., equilibria with time-independent flows, are not Lagrangian equilibria.   To accommodate stationary equilibria,  the description developed in Sec.~\ref{ssec:ghrcs} in terms  of the relabeled  Lagrangian  variables $\bfQ$ and $\bfPi$ is convenient.

Consider what happens to the Hamiltonian formalism in terms of $(\bfQ,\bfPi)$ when an expansion about a given time-dependent reference trajectory is effected as follows:
\bq
\bfQ= \bfQ_r(\bfb,t) + \bfet(\bfb,t)\,,
\quad
\bfPi= \bfPi_r(\bfb,t) + \bfpi_{\eta}(\bfb,t)\,,
\eq
where $\bfet$ and $\bfpi_{\eta}$ will eventually be seen to be related to  the displacements in the linear energy principles.  But, for now, $ \bfQ_r(\bfb,t)$ and $\bfPi_r(\bfb,t)$  are completely arbitrary.   Expanding
\bq
\p_t\bfQ=\frac{\de \tilde H}{\de \bfPi} \qquad {\rm and} \qquad \p_t\bfPi=- \frac{\de \tilde H}{\de \bfQ}
\eq
about the reference trajectory gives the leading order equations
\bqy
\p_t\bfQ_r&=&\frac{\bfPi_r}{\tilde \rho} -\bfV \cdot\nabla_b \bfQ_r\,,
\nonumber\\
\p_t\bfPi_r&=&- \nabla_b\cdot\left( \bfV \otimes \bfPi_r\right) +  \, \bfF_r\,,
\label{reftraj}
\eqy
where $\bfF_r$ comes from the $W$ part of the Hamiltonian. Now,  it is  supposed  that $(\bfQ_r,\bfPi_r)$ is an  equilibrium state, meaning $\p_t\bfQ_r= \p_t\bfPi_r=0$,
whence from (\ref{reftraj}) if follows that
\bq
\nabla_b\cdot(\tilde{\rho}\,  \bfV \bfV\cdot\nabla_b \bfQ_e)=   \bfF_e\,,
\label{equil}
\eq
where the subscript $r$ has been replaced by $e$ to indicate  an  equilibrium state.
Note, that in this expression $\bfV$ and $\tilde\rho$ could depend on time and we could add explicit time dependence to $W$ to produce a  moving state with the balance of (\ref{equil}).   However, only static Lagrangian equilibria where $\bfV(\bfb)$, $\tilde\rho(\bfb)$,  $\bfQ_e(\bfb)$, and $\bfPi_e(\bfb)$ are considered.  And, most importantly, it is assumed that $\bfQ_e(\bfb)=\bfb$.
{\it With this choice,  static equilibria in the present context correspond to a moving state in the original $(\bfq,\bfpi)$ context,  which in turn will be seen to correspond to stationary  equilibria in the Eulerian context.}  At this point  the relabeling is   connected to the dynamics:  up to now it has been arbitrary. To see this,  return to what this all means in terms of the variable $\bfq$\,:
$
\bfb=\bfQ_e(\bfb)= \bfq_e(\mfA_e(\bfb,t),t)=\mfB_e(\bfa,t)
$.
From the definition of $\bfV$ of (\ref{VR})
$
\bfV(\bfb,t)=\dot\mfB_e(\mfA_e(\bfb,t),t)=\bfv_e(\bfb)
$,
where $\bfv_e(\bfb)$  denotes an Eulerian equilibrium state.  Upon setting $\bfb=\bfx$, i.e., identifying the Eulerian observation point with the moving label,    (\ref{equil}) becomes the usual stationary equilibrium equation,
\bq
\nabla\cdot(\rho_e \bfv_e \bfv_e)=  \bfF_e\,,
\eq
and $\tilde\rho$ becomes the usual  equilibrium $\rho_e(\bfx)$.  It can be shown that  $\bfv_e\cdot \na s_e=0$, $ \na\cdot (\rho_e \bfv_e)=0$, and $\bfv_e\cdot \nabla \bfB_e - \bfB_e\cdot \nabla \bfv_e  +  \bfB_e \nabla\cdot \bfv_e=0$
 follow from (\ref{tildes}), (\ref{tildeden}), and (\ref{tildeB}), respectively.

Next  an expansion  about a stationary Eulerian equilibrium, which in this context is a static Lagrangian equilibrium, can be effected to obtain a quadratic energy functional that has no explicit time dependence.   The identification of $\bfb=\bfx$ leads  to a usual procedure of measuring perturbed quantities  relative to the unperturbed trajectories of a stationary equilibrium, as in \cite{Frieman1960,morrison98} for fluid models  (and \cite{Morrison1989,Morrison1990} for kinetic theories).  However, here, evidently for the first time,  this idea has been incorporated on the nonlinear level before expansion and treated in the purely Hamiltonian framework.

Before considering stationary equilibria,   the usual `$\delta W$' energy principle for static equilibrium of  Refs.~\cite{Bernstein1958b,lust-shluter,hain} will be treated.   This principle is an infinite-dimensional version of Lagrange's necessary and sufficient stability theorem of mechanics (see, e.g.,  \cite{wintner}),  which is applicable to natural Hamiltonians of the separable form, kinetic plus potential.
For static MHD configurations the relabeling of Sec.~\ref{ssec:ghrcs} is not necessary and the variables $(\bfq,\bfpi)$ are sufficient; in fact, $\bfQ=\bfq$,  $\bfPi=\bfpi$, $\bfV\equiv 0$; thus,  the equilibrium is described by
\bq
\bfb=\bfQ_e(\bfb)=\bfq_e(\bfa)=\bfa=\bfx\,.
\label{statice}
\eq
Since static  equilibria are given by time-independent $\mathbf{q}_{e}$, this point can be taken to be the Eulerian observation position, i.e., \ $\bfq_e=\bfa=\bfx$ as given in  (\ref{statice}).
Since the Hamiltonian (\ref{eq:H_Lagr}) is of separable form, Lagrange's theorem would imply that the equilibrium is stable if and only if $\mathbf{q}_{e}$ is a local minimum of the potential energy,%
\begin{equation}
W[\mathbf{q}]=\inta \left[  \rho_{0}U\left(  s_{0},\rho_{0}/J\right)
+ \frac{\partial q_i}{\partial a^{j}}\frac{\partial q^{i}%
}{\partial a^{k}}
 \frac{B_{0}^{j}B_{0}^{k}}{8\pi J}
 \right] \,.
 \label{eq:W_Lagr}%
\end{equation}
There are mathematical subtleties to this theorem, even in the finite-dimensional case, but as is common in plasma physics this formal statement of Lagrange's theorem will be assumed.
Following convention,  the infinitesimal displacement from static equilibria $\bfq_e$ will be denoted by
$\boldsymbol{\xi}$, a displacement relative to an inertial frame,   instead of $\bfet$,  i.e., $\mathbf{q}=\mathbf{q}_{e}+\boldsymbol{\xi}$.
The second variation of the potential energy (\ref{eq:W_Lagr}) (see \cite{newcomb} and
\cite{morrison98} for details)   gives
\begin{align}
\delta^{2}W\left[ Z_{e};\boldsymbol{\xi}\right]   &  =\frac{1}%
{2}\intx \Big[  \left(  \rho_{e}\frac{\partial p_{e}}{\partial\rho_{e}%
} +\frac{B_{e}^{2}}{4\pi}\right)  \left(  \partial_{i}\mathbf{\xi}^{i}\right)
^{2}   \label{eq:dW_New}\\
 &+ \left( p_{e}+\frac{B_{e}^{2}}{8\pi}\right)  \left(  \left(
\partial_{i}\mathbf{\xi}^{j}\right)  \left(  \partial_{j}\mathbf{\xi}%
^{i}\right)  -\left(  \partial_{i}\mathbf{\xi}^{i}\right)  ^{2}\right) \nonumber\\
&   + \left[  \left(  \partial_{j}\mathbf{\xi
}_{i}\right)  \left(  \partial_{k}\mathbf{\xi}^{i}\right)  -2\left(
\partial_{i}\mathbf{\xi}^{i}\right)  \left(  \partial_{k}\mathbf{\xi}%
^{j}\right)  \Big]
\frac{B_{e}^jB_{e}^k}{4\pi}
 \right] \, ,
\nonumber
\end{align}
where $\partial_{i}\mathbf{\xi}^{j}:=\partial\xi^{j}/\partial q_{e}^{i}=\partial\xi^{j}/\partial {x}^i$  and
where the Eulerian static equilibrium quantities, denoted by $e$,  are consistent with  the Grad-Shafranov equation (see \cite{amp1}).  Particular care should be paid to  the treatment of boundary  terms  as pointed out in Ref.~\cite{newcomb}  but, as noted above,  an in-depth treatment of boundary conditions, including the plasma vacuum interface will be considered elsewhere.

Given an equilibrium solution, the functional $\delta^{2}W$ is typically
viewed as a quadratic form in  $\boldsymbol{\xi}$, viz.\  a functional that, upon
variation, defines the linear dynamics of perturbations with respects to the
equilibrium. By carrying out some manipulations, the functional
(\ref{eq:dW_New}) can be transformed in the more familiar expression of Ref.~\cite{Bernstein1958b}, %
\bqy
\delta^{2}W\left[ Z_e;  \boldsymbol{\xi}\right]  &=&\frac{1}{2}\!\intx\!
 \Big[
\rho_{e}\frac{\partial p_{e}}{\partial\rho_{e}}\left(  \mathbf{\nabla}%
\cdot\boldsymbol{\xi}\right)^{2}
+ \left(  \mathbf{\nabla}\cdot
\boldsymbol{\xi }\right)  \left(  \mathbf{\nabla}p_{e}\cdot
\boldsymbol{\xi}\right)
\nonumber\\
&{\ }&\quad + \frac{\left\vert \de \bfB \right\vert^{2} }{4\pi}
+ \mathbf{j}_{e}\times\boldsymbol{\xi}\cdot\de \bfB
\Big] \,,
\label{eq:dW_Ber}%
\eqy
where $4\pi \mathbf{j}%
_{e}= \mathbf{\nabla}\times\mathbf{B}_{e}$ is the equilibrium
current and $\de \bfB :=\mathbf{\nabla}\times\left(\boldsymbol{\xi}\times
\mathbf{B}_{e}\right)$.
The linear Hamiltonian is given by
\bq
H_{\rm stc}[\boldsymbol{\xi}, \bfpi_{\xi}]=\de^2H= \intx \frac{|\bfpi_{\xi}|^2}{2\rho_e} +\de^2 W\,,
\label{FRd2H}
\eq
which with the linear Poisson bracket
\bq
\{F,G\}_{(\boldsymbol{\xi},\bfpi_{\xi})}:=\intx\left(
\frac{\de F}{\de \boldsymbol{\xi}} \cdot \frac{\de G}{\p \bfpi_{\xi}}
-
\frac{\de G}{\de \boldsymbol{\xi}} \cdot \frac{\de F}{\p\bfpi_{\xi}}
\right)\\,
\eq
produces the linear Hamiltonian system, obtained by expansion about static equilibria, as
\bqy
\dot{\boldsymbol{\xi}}&=&\{{\boldsymbol{\xi}},H_{\rm stc}\}_{(\boldsymbol{\xi},\bfpi_{\xi})}= \frac{\de H_{\rm stc}}{\de \bfpi_{\xi}}
\nonumber\\
{\dot{\bfpi}}_{\xi}&=&\{\bfpi,H_{\rm stc}\}_{(\boldsymbol{\xi},\bfpi_{\xi})}= -\frac{\de H_{\rm stc}}{\de \boldsymbol{\xi}} \,.
\eqy
Thus, this Hamiltonian system is considered to be stable by Lagrange's theorem if and only if $\delta^{2}%
W$\ is positive for any perturbation $\boldsymbol{\xi}$, i.e. if and only if
the quadratic form is positive definite.

Now consider Eulerian stationary equilibria in the Lagrangian variable framework,  using the relabeling transformation discussed above and  in Sec.~\ref{ssec:ghrcs}.  For such equilibria,  Lagrange's theorem in general does not apply:  because of the presence of  the fictitious term,  the Hamiltonian is no longer of separable form and instead  one only  has  Dirichlet's sufficient condition for stability.  For stationary equilibria the analog of (\ref{statice}) is
$
\bfb=\bfQ_e(\bfb)=\bfq_e(\mfA(\bfb,t),t)=\bfx
$,
the displacement relative to the relabeled fluid elements is given by
$
\bfet(\bfx,t)=\left.\bfxi(\bfa,t)\right|_{\bfa=\bfq_e^{-1}(\bfx,t)}
$,
and stationary equilibrium quantities are represented in terms of unrelabeled fluid elements by
\begin{align}
\rho_{e}\left(  \mathbf{x}\right) &=
\left.  \frac{\rho_{0}\left(  \mathbf{a}\right)  }{J\left(  \mathbf{a}%
,t\right)  }\right\vert _{\mathbf{a}=\mathbf{q}_{e}^{-1}\left(  \mathbf{x}%
,t\right)  }\,,\\
v_{ei}\left(  \mathbf{x}\right) &=
\left.  \frac{\pi_{i}\left(  \mathbf{a},t\right)  }{\rho_{0}\left(
\mathbf{a}\right)  }\right\vert _{\mathbf{a}=\mathbf{q}_{e}^{-1}\left(
\mathbf{x},t\right)  }\, ,\\
s_{e}\left(
\mathbf{x}\right) &=
\left.  s_{0}\left(  \mathbf{a}\right)  \right\vert _{\mathbf{a}%
=\mathbf{q}_{e}^{-1}\left(  \mathbf{x},t\right)  } \, ,\\
B^i_{e}\left(  \mathbf{x}\right) &=
\left.  \frac{\partial q^{i}\left(  \mathbf{a},t\right)  }{\partial a^{j}%
}\frac{B^j_{0}\left(  \mathbf{a}\right)  }{J\left(  \mathbf{a},t\right)
}\right\vert_{\mathbf{a}=\mathbf{q}_{e}^{-1}\left(  \mathbf{x},t\right)  }
 \,.
\end{align}
Following Ref.~\cite{morrison98},  the second variation of the
Hamiltonian  in terms of the canonically conjugate variables $(\boldsymbol{\eta},\bfpi_{\eta})$ results%
\bqy
\delta^{2}H_{\mathrm{la}}\left[ Z_e; \boldsymbol{\eta},\boldsymbol{\pi}_{\eta}\right]  &=&
\frac{1}{2}\intx \Big[
  \frac{1}{\rho_e}\big|  \boldsymbol{\pi}_{\eta}%
-\rho_e\mathbf{v}_e\cdot\mathbf{\nabla}\boldsymbol{\eta}\big|^{2}%
\nonumber\\
&{\ }& \hspace{.75 in} + \boldsymbol{\eta}\cdot\mathfrak{V}_e\cdot\boldsymbol{\eta}
\Big]
\label{eq:d2H_Lagr}%
\eqy
where $\rho_e$, $\mathbf{v}_e$,  and the operator $\mathfrak{V}_e$\ have
no explicit time dependence. The functional
\bqy
\delta^{2}W_{\mathrm{la}}\left[  \boldsymbol{\eta}\right]  &:=&\frac{1}{2}%
\intx   \boldsymbol{\eta}\cdot\mathfrak{V}_e\cdot
\boldsymbol{\eta}
\nonumber\\
&=&\frac{1}{2}\intx \Big[  \rho_e\left(
\mathbf{v}_e\cdot\nabla\mathbf{v}_e\right)  \cdot\left(  \boldsymbol{\eta}\cdot \nabla
\boldsymbol{\eta}\right)
\nonumber\\
&{\ }&
\hspace{.2 in} -\rho_e\left|  \mathbf{v}_e%
\cdot\nabla\boldsymbol{\eta}\right|^{2}\Big]
+\delta^{2}W\left[
\boldsymbol{\eta}\right]\,,
\eqy
which is that obtained  by Frieman and Rotenberg in Ref.~\cite{Frieman1960}, represents a generalization to stationary equilibria of the potential energy of  (\ref{eq:dW_Ber}).  The linear Hamiltonian about stationary equilibria is  $H_{\rm str}=\de^2H_{\mathrm{la}}$ with (\ref{eq:d2H_Lagr}), and the linear equations of motion are
\bqy
\frac{\partial\boldsymbol{\eta}}{\partial t}&=&
\{\bfet,H_{\rm str}\}_{(\bfet,\bfpi_{\eta})}= \frac{1}{\rho_e}
\left(
\boldsymbol{\pi}_{\eta}-\rho_e\mathbf{v}_e\cdot {\nabla}%
\boldsymbol{\eta }
\right)
 \label{eq:d_eta}\\
\frac{\partial\boldsymbol{\pi}_{\eta}}{\partial t}&=&\{\bfpi_{\eta},H_{\rm str}\}_{(\bfet,\bfpi_{\eta})}
= -\rho_e\mathbf{v}_e%
\cdot\mathbf{\nabla}\frac{\boldsymbol{\pi}_{\eta}}{\rho_e}
\nonumber\\
&+& \mathbf{\nabla}
\cdot\left(  \rho_e\boldsymbol{\eta}\mathbf{v}_e\cdot\mathbf{\nabla v}_e\right)
 +{\nabla}\left(  \rho_e\frac{\partial p_e}{\partial\rho_e}\mathbf{\nabla
}\cdot\boldsymbol{\eta}+\boldsymbol{\eta}\cdot\mathbf{\nabla}p_e\right)
\nonumber\\
&+&\frac{1}{4\pi}\left[  \mathbf{B}_e\cdot{\nabla \de\bfB}+{\de \bfB}%
\cdot \nabla \bfB_e - {\nabla}\left(  \mathbf{B}_e\cdot {\de\bfB}%
\right)  \right]  ,\label{eq:d_pi_eta}%
\eqy
where  the equilibrium equations have been used to simplify the functional derivative
of  $H_{\rm str}$ with respect to $\boldsymbol{\eta}$. By exploiting
the relation
\bqy
\frac{\partial\boldsymbol{\pi}_{\eta}}{\partial t}+\rho_e\mathbf{v}%
\cdot {\nabla}\frac{\boldsymbol{\pi}_{\eta}}{\rho_e}&=&\rho_e\frac
{\partial^{2}\boldsymbol{\eta}}{\partial t^{2}} + 2\rho_e\mathbf{v}_e\cdot
 {\nabla}\frac{\partial\boldsymbol{\eta}}{\partial t}
 \nonumber\\
 &{\ }&
 + {\nabla
}\cdot\left(  \rho_e\mathbf{v}_e\bfv_e\cdot {\nabla}\boldsymbol{\eta}\right)\,,
\eqy
Eqs.~(\ref{eq:d_eta}) and (\ref{eq:d_pi_eta}) can be put into  the
form of  \cite{Frieman1960}, i.e. as%
\begin{equation}
\rho_e\frac{\partial^{2}\boldsymbol{\eta}}{\partial t^{2}}
+2\rho_e\mathbf{v}_e%
\cdot {\nabla}\frac{\partial\boldsymbol{\eta}}{\partial t}-\mathbf{F}^v_e=0,
\end{equation}
where the force operator with velocity terms is
\bqy
\mathbf{F}^v_e\left(  \boldsymbol{\eta}\right)  &:=& {\nabla}\cdot\left(
\rho_e\boldsymbol{\eta}\mathbf{v}_e\cdot \nabla \bfv_e-\rho_e\mathbf{v}_e\bfv_e%
\cdot {\nabla}\boldsymbol{\eta}\right)  \nonumber\\
&{\  }& \hspace{.25 cm} + {\nabla}\left(
\rho_e\frac{\partial p_e}{\partial\rho_e} {\nabla}\cdot
\boldsymbol{\eta }
+\boldsymbol{\eta}\cdot {\nabla}p_e\right)
\nonumber\\
&{\ \,}& \hspace{.25 cm} +
\frac{1}{4\pi}\left[  \mathbf{B}_e\cdot \nabla \de\bfB + \de\bfB\cdot
\mathbf{\nabla B}_e- {\nabla}\left(  \mathbf{B}_e\cdot \de\bfB\right)
\right] \, .
\nonumber
\eqy
In this case, it is clear from the expression (\ref{eq:d2H_Lagr}) that the
Hamiltonian is not of separable form and in general Lagrange's theorem does
not apply. However, due  to the arbitrariness of $\boldsymbol{\pi}_{\eta}$, which does not
contributes to $\delta^{2}W_{\mathrm{la}}$, the quadratic term in the
integrand of Eq. (\ref{eq:d2H_Lagr}) can be put equal to zero and a sufficient
condition for stability is given by $\delta^{2}W_{\mathrm{la}}>0$  for any
perturbation $\boldsymbol{\eta}$.  This is an infinite-dimensional  version of Dirichlet's theorem.

For completeness  we record the first order Eulerian perturbations that are induced by the 
Lagrangian variation $\boldsymbol{\xi}$, written in terms of the `Eulerianized' 
displacement  $\boldsymbol{\eta}$.  They are given as follows: 
\begin{align}
\delta\rho_{\mathrm{la}} &  =- {\nabla}\cdot\left( \rho_{e}\boldsymbol{\eta}\right)
\label{eq:laDrho} \\
\delta\mathbf{M}_{\mathrm{la}} &  =\boldsymbol{\pi}_{\eta}-\rho_e
\boldsymbol{\eta}\cdot\mathbf{\nabla v_e-v_e\nabla}\cdot\left(  \rho_e
\boldsymbol{\eta}\right)
\label{eq:laDM}
\\%
\delta {\si}_{\mathrm{la}}&=-  {\nabla}\cdot(\si_e \boldsymbol{\eta})
\label{eq:laDs}%
\\
\delta\mathbf{B}_{\mathrm{la}}&= - {\nabla}\times
\left(  \mathbf{B}_{e}   \times   \boldsymbol{\eta}
\right)
\label{eq:laDB}%
\end{align}
where the momentum and entropy perturbations $\delta \mathbf{M}_{\mathrm{la}}$ 
and $\delta {\si}_{\mathrm{la}}$ can be replaced by the following velocity and  pressure perturbations:
\begin{eqnarray}
\delta p_{\mathrm{la}} &=&-\gamma p_e {\nabla}\cdot\boldsymbol{\eta} 
-\boldsymbol{\eta}\cdot\mathbf{\nabla}p_e
\\
\delta\mathbf{v}_{\mathrm{la}}  &=&\frac{\partial\boldsymbol{\eta}}{\partial t}+\mathbf{v}_e%
\cdot \nabla\boldsymbol{\eta}-\boldsymbol{\eta}\cdot\nabla \mathbf{v}_e \label{eq:laDv}
\end{eqnarray}
as were used in Ref.~\cite{Frieman1960}. 


\section{Eulerian stability -- the energy-Casimir method}
\label{sec:ec}

Before proceeding we briefly list results of Ref. \cite{amp1} for readability
and self-containness.

The Hamilton form of the MHD equations in the Eulerian variables $Z=\left(
\rho,\mathbf{v},s,\mathbf{B}\right)  $ is
\begin{equation}
\frac{\partial Z}{\partial t}=\left\{  Z,H\right\}  \,,
\label{eq:genEOM}%
\end{equation}
where $H[Z]$, the Hamiltonian is the energy expressing of Eq.~(\ref{eq:H_Euler}),
and $\left\{  \cdot,\cdot\right\}  $ represents the noncanonical Poisson
bracket of \cite{morrison-greene,morrison-greene-cor}, which is not of canonical form because the
Eulerian variables $Z$ are not canonical variables.

Because Eulerian variables are not canonical variables, the noncanonical
Poisson bracket has degeneracy that gives rise to Casimir invariants, special
invariants $C$ that satisfy $\{C,F\}=0$ for all functionals $F$, and these
give rise to the variational principles of \cite{amp1} that we can use further
for determining stability. This follows from the general form for noncanonical
Poisson brackets given by
\begin{equation}
\left\{  F,G\right\}  =\intx \frac{\delta F}{\delta Z}\cdot\mathbb{J}%
\cdot\frac{\delta G}{\delta Z}\, \label{ncbkt}%
\end{equation}
which is defined on two functionals $F$ and $G$. Here $\mathbb{J}$, the
cosymplectic operator, is formally anti-self-adjoint and must satisfy a
strenuous condition for the Jacobi identity \cite{morrison82,morrison98}. In
terms of (\ref{ncbkt}) the equilibrium variational principles of \cite{amp1}
amount to $\delta\mathfrak{F}/\delta Z=0\Rightarrow\dot{Z}=0$, where
$\mathfrak{F}:=H+C$. This follows from
$
\dot{Z}=\mathbb{J}\cdot\frac{\delta H}{\delta Z}=\mathbb{J}\cdot\frac
{\delta\mathfrak{F}}{\delta Z}=0
$
since $\mathbb{J}\cdot\delta C/\delta Z=0$ by definition. In
finite-dimensional Hamiltonian systems Dirichlet \cite{lagrange} showed that
the Hamiltonian provides a sufficient condition for nonlinear stability of an
equilibrium point if energy surfaces in the vicinity of the equilibrium are
ellipsoidal, which is equivalent to definiteness of the second variation of
the Hamiltonian. Carrying this idea over to the present infinite-dimensional
noncanonical setting this amounts to definiteness of the second variation of
$\mathfrak{F}$. This is sufficient for linear stability and points toward
nonlinear stability, but a rigorous mathematical proof requires information
about the existence of solutions for MHD which is a famous open problem. It
should be noted that by nonlinear stability we mean stability to infinitesimal
perturbations under the full nonlinear dynamics of the system. A nonlinearly
unstable system is unstable to infinitesimal perturbations as opposed to
finite amplitude instability which requires a sufficiently large perturbation
for instability as with a dimpled potential well.

In this section we consider MHD with symmetry, as described in \cite{amp1}.
All geometric symmetries can be described as a combination of axial symmetry
and translational symmetry or, in other terms, as a helical symmetry. Given a
cylindrical coordinate system $\left(r,\phi,z\right)$, we define a helical
coordinate $u=\phi\left[  l\right]  \sin\alpha+z\cos\alpha$,
where $\left[l\right]$  is a scale length and $\alpha$\ defines the
helical angle. The unit vector in the direction of \ the coordinate $u$\ can
be written as $
\mathbf{u}=kr\nabla u=\left(  k\left[  l\right]  \sin\alpha\right)
\mathbf{\phi}+\left(  kr\cos\alpha\right)  \mathbf{z}$,  
where $k^2=1/\big(  \left[  l\right]  ^{2}\sin^{2}\alpha+r^{2}\cos^{2}
\alpha\big)$ represents a metric factor.  Thus the second helical
direction results,   
$
\mathbf{h=}kr\, \mathbf{\nabla}r\times\mathbf{\nabla}u=-\left(  kr\cos
\alpha\right)  \mathbf{\phi}+\left(  k\left[  l\right]  \sin\alpha\right)
\mathbf{z}%
$, 
and the helical symmetry is expressed by the fact that 
$
\mathbf{h}\cdot\mathbf{\nabla}f=0,
$
where $f$ is a generic scalar function.  The direction $\mathbf{h}$, which is
called the symmetry direction, can be chosen to obtain axial ($\alpha=0$),
translational ($\alpha=\pi/2$),  or {true} helical ($0<\alpha<\pi/2$) symmetry
and the metric factor $k$ changes accordingly. {In the following we use the
identities} 
\begin{equation}
\mathbf{\nabla}\cdot\mathbf{h}=0\,, \qquad\mathbf{\nabla}\times\left(
k\mathbf{h}\right)  =-\left(  k^{3}\left[  l\right]  \sin2\alpha\right)
\mathbf{h}\,, 
\end{equation}
{which also show that, for }$\sin2\alpha=0${, we can define a coordinate in
the symmetry direction as }$\mathbf{\nabla}h=k\mathbf{h}${. }

Using the notation described before, the magnetic field can be rewritten as 
\begin{equation}
\mathbf{B}\left(  r,u\right)  =B_{h}\left(  r,u\right)  \mathbf{h}%
+\mathbf{\nabla}\psi\times k\mathbf{h}\,, 
\label{symB}
\end{equation}
where $\psi=\psi\left(  r,u\right)  $\ is the magnetic flux function, while   
the velocity becomes  
\begin{equation}
\mathbf{v}\left(  r,u\right)  =v_{h}\left(  r,u\right)  \mathbf{h}%
+\mathbf{v}_{\bot}\left(  r,u\right)  \label{eq:Mhel}%
\end{equation}

In terms of the variables $Z_{S}=\left(  \rho,\mathbf{v}_{\bot},v_{h}%
,\psi,B_{h}\right)  $, where we assume the entropy is a flux function, i.e.
$s=\mathcal{S}\left(  \psi\right)  $, the energy-Casimir functional is given
by
\begin{align}
\mathfrak{F}  &  =%
\intx
\Big(  \frac{\rho |\mathbf{v}_{\bot}|^{2}}{2}+\frac{\rho v_{h}^{2}}{2}+\rho
U+\frac{k^{2}\left\vert \nabla\psi\right\vert ^{2}}{8\pi}+\frac{B_{h}^{2}%
}{8\pi}  \nonumber\\
& \quad -    \rho\mathcal{J}-kB_{h}\mathcal{H}-\left(  k^{4}\left[  l\right]
\sin2\alpha\right)  \mathcal{H}^{-}
\nonumber\\
&\qquad -\frac{\rho}{k}  v_{h}\mathcal{G}%
-\mathbf{v}\cdot\mathbf{B}\,\mathcal{F}\Big) \,. \label{hpc}%
\end{align}
where $\mathcal{F}$,$\ \mathcal{G}$, $\mathcal{H}$ and $\mathcal{J}$ are four
arbitrary functions of $\psi$ and $\mathcal{H}^{-}(\psi):=\int^{\psi
}\mathcal{H}\left(  \psi^{\prime}\right)  \,d\psi^{\prime}$.

As discussed in Ref. \cite{amp1}, the first variation of (\ref{hpc}) gives the
equilibrium equations%
\bqy
0&=&\rho\mathbf{v}-\mathcal{F}\mathbf{B}\,-\frac{1}{k}\rho\mathcal{G}\mathbf{h}
   \label{eq:eq2}\\
0&=&\frac{|\mathbf{v}|^{2}}{2}+U+\frac{p}{\rho}-\mathcal{J}-\frac{1}{k}v_{h}\mathcal{G},
   \label{eq:eq3}\\
0&=&\frac{B_{h}}{4\pi}-k\mathcal{H}-v_{h}\mathcal{F}  ,
\label{eq:eq4}\\
0&=& -\mathbf{\nabla}\cdot\left(  \frac{k^{2}}{4\pi}\mathbf{\nabla}\psi\right)
+\rho T\mathcal{S}^{\prime}-\rho\mathcal{J}^{\prime}-kB_{h}\mathcal{H}%
^{\prime}
 \nonumber\\
&{\ }&
 \hspace{1 cm}
 -k^{4}\left[  l\right]  \sin2\alpha\,\mathcal{H}
 -\frac{1}{k}\rho v_{h}\mathcal{G}^{\prime}-\mathbf{v\cdot B}\,\mathcal{F}%
^{\prime}
 \nonumber\\
&{\ }&
 \hspace{2 cm}
 +\nabla\cdot\left(  \mathcal{F}k\mathbf{h}\times\mathbf{v}_{\bot
}\right).
\eqy
where in the last equation primes indicate derivatives with respect to $\psi$.
We assume the equilibrium is known and proceed to the second variation.

The second variation of the energy-Casimir principle results%
\begin{equation}
\delta^{2}\mathfrak{F}=%
\intx
\left(  \delta\tilde{Z}_{S}\cdot\mathbb{K}\cdot\delta\tilde{Z}_{S}^{^{t}%
}\right)    \label{eq:d2F}%
\end{equation}
where we rearranged the terms in the vectorial form%
\begin{equation}
\delta\tilde{Z}_{S}=%
(
\delta v_{r},   \delta v_{u},   \delta v_{h},   k\partial_{r}\delta\psi,
k\partial_{u}\delta\psi,  \delta B_{h},  \delta\rho,  \delta\psi)
\nonumber
\end{equation}
and the matrix quadratic form%
 \begin{widetext}
\begin{equation}
\mathbb{K}=%
\begin{bmatrix}
\rho & 0 & 0 & 0 & -\mathcal{F} & 0 & \frac{1}{\rho}\mathcal{F}B_{r} &
-\mathcal{F}^{\prime}B_{r}\\
^{\prime\prime} & \rho & 0 & \mathcal{F} & 0 & 0 & \frac{1}{\rho}%
\mathcal{F}B_{u} & -\mathcal{F}^{\prime}B_{u}\\
\qquad & ^{\prime\prime} & \rho & 0 & 0 & -\mathcal{F} & \frac{1}{\rho
}\mathcal{F}B_{h} & -\mathcal{F}^{\prime}B_{h}\,-\frac{1}{k}\rho
\mathcal{G}^{\prime}\\
& \qquad & ^{\prime\prime} & \frac{1}{4\pi} & 0 & 0 & 0 & \frac{1}{\rho
}\mathcal{FF}^{\prime}B_{u}\\
&  & \qquad & ^{\prime\prime} & \frac{1}{4\pi} & 0 & 0 & -\frac{1}{\rho
}\mathcal{FF}^{\prime}B_{r}\\
&  &  & \qquad & ^{\prime\prime} & \frac{1}{4\pi} & 0 & -\frac{1}{\rho
}\mathcal{FF}^{\prime}B_{h}-k\mathcal{H}^{\prime}-\frac{1}{k}\mathcal{G}%
\,\mathcal{F}^{\prime}\\
&  &  &  & \qquad & ^{\prime\prime} & \frac{1}{\rho}c_{s}^{2} & -\frac
{1}{k\rho}\left(  B_{h}\,\mathcal{F}+\frac{1}{k}\rho\mathcal{G}\right)
\mathcal{G}^{\prime}-\mathcal{J}^{\prime}\\
&  &  &  &  & \qquad & ^{\prime\prime} & -\Upsilon
\end{bmatrix}
, \label{eq:K}%
\end{equation}
\end{widetext}
Since $\mathbb{K}$\ is symmetric, we only showed the terms above and on the
diagonal. The notation $\delta\tilde{Z}_{S}$ is used to emphasize that in the
rearranged vectorial form we have included, as separated elements, not only
the perturbations of the variables $Z_{S}$ but also the spatial derivatives of
the perturbation $\delta\psi$. Moreover, in the quadratic form (\ref{eq:K}),
the notation $c_{s}^{2}=\partial p/\partial\rho$\ represents the square of the
plasma sound speed, while $\Upsilon$\ indicates the expression%
\bqy
\Upsilon&=&-\rho T\mathcal{S}^{\prime\prime}-\rho\left(  \partial T/\partial
s\right)  \mathcal{S}^{\prime2}+\rho\mathcal{J}^{\prime\prime} + kB_{h}%
\mathcal{H}^{\prime\prime}
\nonumber\\
&+&\left(  k^{4}\left[  l\right]  \sin2\alpha\right)
\mathcal{H}^{\prime}+\frac{1}{k}B_{h}\mathcal{FG}^{\prime\prime}+\frac
{1}{k^{2}}\rho\mathcal{GG}^{\prime\prime}
\nonumber\\
&+&\frac{B^{2}}{\rho}\,\mathcal{FF}%
^{\prime\prime}+B_{h}\frac{1}{k}\mathcal{GF}^{\prime\prime}
\,.%
\eqy

Next, we consider Eq. (\ref{eq:eq2}) and we define%
\begin{equation}
\mathbf{S}:=\rho\mathbf{v}-\mathcal{F}\mathbf{B}-\frac{1}{k}\rho
\mathcal{G}\mathbf{h}\,. \label{eq:nP}%
\end{equation}
At the equilibrium $\mathbf{S}=0$ and we use this equation to obtain
$\mathbf{v}$ as a function of the other variables. Next, we define%
\begin{equation}
Q:=\frac{B^{2}}{2\rho^{2}}\mathcal{F}^{2}+U+\frac{p}{\rho}-\mathcal{J}%
-\frac{1}{2k^{2}}\mathcal{G}^{2}, \label{eq:nQ}%
\end{equation}
which follows from Eq. (\ref{eq:eq3}) after substitution of Eq. (\ref{eq:eq2}%
). Again, $Q=0$ at the equilibrium and we can use this equation to obtain
$\rho=\rho\left(  \psi,\mathbf{B}\right)  $. Last, we define%
\begin{equation}
\mathbf{R}:=\frac{\left(  1-\mathrm{M}^{2}\right)  }{4\pi}\mathbf{B}-\left(
k\mathcal{H}+\frac{1}{k}\mathcal{FG}\right)  \mathbf{h} \label{eq:nR}%
\end{equation}
where, substituting the previous results, the poloidal Alfv\'en Mach number%
\begin{equation}
\mathrm{M}^{2}=\frac{4\pi\rho |\mathbf{v}_{\bot}|^{2}}{B_{\bot}^{2}}=\frac{4\pi
\mathcal{F}^{2}}{\rho\left(  \psi,\mathbf{B}\right)  }%
\end{equation}
is considered as a function of $\psi$\ and $\mathbf{B}$. The component of
$\mathbf{R}$ in the symmetry direction is related to Eq. (\ref{eq:eq4}) (after
substituting of $v_{h}$ and $\rho$) and at the equilibrium $R_{h}=0$, whereas
$\mathbf{R}_{\bot}$\ is related to the derivatives of $\psi$\ and in general
is not zero.

By introducing in Eq. (\ref{eq:d2F}) the variations%
\begin{align}
\delta\mathbf{S}  &  =\rho\delta\mathbf{v}+\mathbf{v}\delta\rho-\mathcal{F}%
\delta\mathbf{B}-\mathcal{F}^{\prime}\mathbf{B}\delta\psi-\frac{\mathbf{h}}{k}\left(
\mathcal{G}\delta\rho+\rho\mathcal{G}^{\prime}\delta\psi\right)
,
\nonumber\\
\delta Q  &  =\frac{1}{\rho}\left(  c_{s}^{2}-\mathrm{M}^{2}c_{a}^{2}\right)
\delta\rho
-\mathcal{M}\, \delta\psi+\frac
{\mathrm{M}^{2}}{4\pi}\frac{\mathbf{B}}{\rho}\delta\mathbf{B},
\nonumber\\
\delta\mathbf{R}  &  =\frac{\left(  c_{s}^{2}+c_{a}^{2}\right)
 \left(
\mathrm{M}_{c}^{2}
-\mathrm{M}^{2}\right)  }{4\pi\left(  c_{s}^{2}%
-\mathrm{M}^{2}c_{a}^{2}\right)  }\delta\mathbf{B}
 \label{eq:varR}\\
& \hspace{.5 in}
- \Big[
 \frac{2\mathcal{FF}^{\prime}}{\rho}\mathbf{B}+ \mathcal{N}\,  \mathbf{h}
%
 +  \frac{\mathrm{M}^{2}\, \mathcal{M} }{\left(  c_{s}^{2}-\mathrm{M}^{2}%
c_{a}^{2}\right)  }\frac{\mathbf{B}}{4\pi} \Big]  \delta\psi, \nonumber%
\end{align}
where
\bqy
\mathcal{M}&:=&\mathcal{GG}^{\prime}/k^2+\mathcal{J}^{\prime}
-\left(  T+\rho\partial T/\partial\rho\right)  \mathcal{S}^{\prime}
- B^{2}\mathcal{FF}^{\prime}/\rho^2\,,
\nonumber\\
\mathcal{N}&:=&   k\mathcal{H}^{\prime}%
+ \mathcal{F}^{\prime}\mathcal{G}/k + \mathcal{FG}^{\prime}/k\,,
\eqy
 $c_{a}^{2}=B^{2}/\left(  4\pi\rho\right)  $ is the square of the Alfv\'en
velocity and $\mathrm{M}_{c}^{2}=c_{s}^{2}/\left(  c_{s}^{2}+c_{a}^{2}\right)
$\ represents the square Alfv\'en Mach number corresponding to the
\textquotedblleft cusp velocity", the second variation of the energy-Casimir
principle can be rewritten in a form for which the matrix $\mathbf{K}%
$\ becomes diagonal, i.e.%
\bqy
\delta^{2}\mathfrak{F}&=&%
\intx
\Big[
a_{1}\left|\delta\mathbf{S}\right|^{2} 
+a_{2}\left(\delta Q\right)^{2}
\nonumber\\
&{\ }&\hspace{.7 in} +a_{3}\left|\delta\mathbf{R}\right|^{2}  
+ a_{4}\left(\delta\psi\right)^{2}
\Big]
 , \label{eq:d2F_diag}%
\eqy
where%
\begin{align}
a_{1}  &  =\frac{1}{\rho},\qquad a_{2}=\frac{\rho}{\left(  c_{s}%
^{2}-\mathrm{M}^{2}c_{a}^{2}\right)  },
\nonumber\\
a_{3}&=\frac{4\pi\left(
c_{s}^{2}-\mathrm{M}^{2}c_{a}^{2}\right)  }{\left(  c_{s}^{2}+c_{a}%
^{2}\right)  \left(  \mathrm{M}_{c}^{2}-\mathrm{M}^{2}\right)  }\,, %
\nonumber\\
a_{4}  &  =-\Upsilon 
-a_{1}\left| \frac{\delta\mathbf{S}}{\delta\psi}\right\vert _{\tilde{Z}_{S}}^{2} 
-a_{2}\left. \frac{\delta Q}{\delta\psi}\right\vert _{\tilde{Z}_{S}}^{2} 
-a_{3}\left|\frac{\delta\mathbf{R}}%
{\delta\psi}\right\vert _{\tilde{Z}_{S}}^{2}. \label{eq:a4}%
\end{align}

{}From Eq.~(\ref{eq:d2F_diag}) it follows that, if $a_{i}>0$\textit{
}for\textit{ }$i=1..4$\textit{.} the equilibrium is a local minimum (i.e.
$\delta^{2}\mathfrak{F}>0$) and stability is thus proved. In particular, the
coefficient $a_{1}$ is always positive while the positiveness of the
coefficients $a_{2}$ and $a_{3}$ can be reduced to the condition
$\mathrm{M}<\mathrm{M}_{c}$. However, the coefficient $a_{4}$ involves in a
very complicated way the second derivatives of the flux functions
$\mathcal{F}$,$\ \mathcal{G}$, $\mathcal{H},$ $\mathcal{J}$ and $\mathcal{S}$
and considerations on its positiveness require, for each specific problem,
specific investigations. Notice that stability is assured for generic Eulerian
perturbations, which in general do not satisfy any of the dynamic constraints
of the equilibrium (e.g. Casimir invariants can be modified by the perturbations).

In the equations above we have considered $\psi$\ and $\mathbf{B}_{\bot}$ as
two independent variables and in the last of Eqs.~(\ref{eq:a4}) the notation $\left.
\delta/\delta\psi\right\vert _{\tilde{Z}_{S}}$ indicates only the terms on the
right hand side of  the first of Eqs.~(\ref{eq:varR}) that multiply
$\delta\psi$. Given an equilibrium, the terms $\left.  \delta\mathbf{S}%
/\delta\psi\right\vert _{\tilde{Z}_{S}}$, $\left.  \delta Q/\delta
\psi\right\vert _{\tilde{Z}_{S}}$, $\left.  \delta\mathbf{R}/\delta
\psi\right\vert _{\tilde{Z}_{S}}$\ are completely known, since they depend
only on equilibrium quantities and not on the perturbations. However, the
variation of the poloidal magnetic field is related to the variation of $\psi$
by the equation
\begin{equation}
\delta\mathbf{B}_{\bot}=\mathbf{\nabla}\delta\psi\times k\mathbf{h}.
\label{eq:var_dB}%
\end{equation}
The stability conditions deduced by considering the positiveness of the
coefficients  of Eqs.~(\ref{eq:a4}) are thus over-estimated and
represents only sufficient criteria.

In order to obtain, for a given equilibrium, a better stability condition, we
exploit the relation Eq.~(\ref{eq:var_dB}) between $\delta\mathbf{B}_{\bot}$ and
$\delta\psi$\ and we consider $\delta^{2}\mathfrak{F}$ as a function of
$\delta\mathbf{S}$, $\delta Q$, $\delta R_{h}$ and $\delta\psi$. Upon
variation with respect to $\delta\mathbf{S}$, $\delta Q$ and $\delta R_{h}$,
it is straightforward to show that the minimum of $\delta^{2}\mathfrak{F}%
$\ corresponds to
\begin{equation}
\delta\mathbf{S}=\delta Q=\delta R_{h}=0, 
\label{eq:min_d2F}%
\end{equation}
provided that $a_{2}$ and $a_{3}$ are positive. Equations (\ref{eq:min_d2F}),
which yield
$
\mathbf{S}=Q=R_{h}={\rm const}=0\,, 
$
correspond to the reduced variational principle presented in Ref.~\cite{amp1},
where the constraints arising from the equilibrium equations have been used to
obtain a variational principle for $\psi$\ alone.

We are reduced to the functional
\begin{equation}
\delta^{2}\mathfrak{F}\left[  \delta\psi\right]  =%
\intx
\left[  a_{3} \left|\delta\mathbf{R}_{\bot}\right|^{2} 
+ a_{4}\left(\delta\psi\right)^{2} \right] , \label{eq:d2F_red}%
\end{equation}
which now depends only on the perturbation $\delta\psi$ and its derivatives.
By using the last expression of (\ref{eq:varR}), which can be rewritten as%
\begin{equation}
\delta\mathbf{R}_{\bot}=\frac{1}{a_{3}}\delta\mathbf{B}_{\bot}+\left.
\frac{\delta\mathbf{R}_{\bot}}{\delta\psi}\right\vert _{\tilde{Z}_{S}}%
\delta\psi
\end{equation}
the second variation of the constrained energy functional becomes%
\bqy
\delta^{2}\mathfrak{F}&=&%
\intx
\bigg[  \frac{1}{a_{3}}\left|\delta\mathbf{B}_{\bot}\right|^{2} 
+2 \left. \frac{\delta\mathbf{R}_{\bot}}{\delta\psi}\right\vert _{\tilde{Z}_{S}}%
\!\!\!\cdot\,  \delta\mathbf{B}_{\bot}\delta\psi
\nonumber\\
&&\hspace{-.8 cm} - \left(  \Upsilon+a_{1}
\left|\frac{\delta\mathbf{S}}{\delta\psi}\right\vert _{\tilde{Z}_{S}}^{2}%
+ a_{2}\left.  \frac{\delta Q}{\delta\psi}\right\vert _{\tilde{Z}_{S}}^{2} 
+a_{3}\left.  \frac{\delta R_{h}}{\delta\psi}\right\vert _{\tilde{Z}_{S}%
}^{2}\right)  \left(  \delta\psi\right)  ^{2}
\bigg]  .
\nonumber
\eqy
Then, we consider the term%
\bqy
2\left.  \frac{\delta\mathbf{R}_{\bot}}{\delta\psi}\right\vert _{\tilde{Z}_{S}}
\!\!\cdot\delta\mathbf{B}_{\bot}\delta\psi
&=&
2 \left.  \frac{\delta \mathbf{R}_{\bot}}{\delta\psi}\right\vert _{\tilde{Z}_{S}}
\!\!\cdot\left(
{\nabla}\delta\psi\times k\mathbf{h}\right)  \delta\psi
\nonumber\\
&=&-\mathbf{\nabla
}\left(  \delta\psi\right)  ^{2}\cdot\left(  \left.  \frac{\delta
\mathbf{R}_{\bot}}{\delta\psi}\right\vert _{\tilde{Z}_{S}}
\!\!\times
k\mathbf{h}\right)
\nonumber ,
\eqy
and, by integrating by parts (neglecting the surface integral), the second
variation becomes%
\begin{equation}
\delta^{2}\mathfrak{F}=%
\intx
\left[  b_{1}\left|\mathbf{\nabla}\delta\psi\right|^{2} + b_{2}\left(
\delta\psi\right)  ^{2}\right] ,
\end{equation}
where $b_{1}={k^{2}}/{a_{3}}$ and
\bqy
 b_{2}&=&-\Upsilon - a_{1}\left\vert  \frac
{\delta\mathbf{S}}{\delta\psi}\right\vert _{\tilde{Z}_{S}}^{2}-a_{2}\left.
\frac{\delta Q}{\delta\psi}\right\vert _{\tilde{Z}_{S}}^{2}-a_{3}\left.
\frac{\delta R_{h}}{\delta\psi}\right\vert _{\tilde{Z}_{S}}^{2}
\nonumber\\
&{\ }&\quad + \mathbf{\nabla
}\cdot\left(  \left. \frac{\delta\mathbf{R}_{\bot}}{\delta\psi}\right\vert
_{\tilde{Z}_{S}}
\!\!\times k\mathbf{h}\right)
\nonumber\,   .
\eqy
Thus, the Euler-Lagrange equation associated with the extrema of
(\ref{eq:d2F_red}) is%
\begin{equation}
\mathbf{\nabla}\cdot\left(  b_{1}\mathbf{\nabla}\delta\psi\right)
-b_{2}\delta\psi=0, \label{eq:d2F_New}%
\end{equation}
which represents a generalized form of the Newcomb equation \cite{newcomb60, goedbloed}
for MHD symmetric equilibria with flow.

Although we have obtained an equation that effectively minimizes the second
variation of the energy-Casimir functional, in most cases the solution of Eq.
(\ref{eq:d2F_New}) requires significative effort. However, a different way to
estimate the minimum of $\delta^{2}\mathfrak{F}$ can be obtained by
introducing into Eq. (\ref{eq:d2F_red}) a Poincar\`{e} inequality
$
\left\Vert \mathbf{\nabla}\delta\psi\right\Vert _{L^{2}}\geq\Lambda\left\Vert
\delta\psi\right\Vert _{L^{2}}%
$,
where $\Lambda$\ is a constant that depends only on the domain and we have
assumed that the mean value of $\delta\psi$\ is zero. A sufficient condition
for the stability results%
\begin{equation}
\delta^{2}\mathfrak{F}\geq%
\intx
\left[  \left(  b_{1\min}\Lambda+b_{2}\right)  \left(  \delta\psi\right)
^{2}\right]
\geq0, \label{eq:d2F_Poin}%
\end{equation}
where $b_{1\min}$\ represents the minimum value of $b_{1}$\ for the considered
equilibrium. The condition (\ref{eq:d2F_Poin}) requires at each point
$
b_{1\min}\Lambda+b_{2}\geq0.
$

\section{Dynamically accessible stability}
\label{sec:da}

Stability of MHD equilibria with flow for perturbations that are
confined to surfaces of constant Casimirs can be assessed by means of the
so-called dynamically accessible variations, which are explicitly constructed
in order to satisfy the Casimir constraints. Dynamically accessible
variations, which are a restricted class of the Eulerian variations presented
in Sec. \ref{sec:ec}, are generated by means of the noncanonical Poisson
bracket of the problem as %
$
\delta Z_{\mathrm{da}}=\left\{  G,Z\right\}
$,
where the functional %
$
G=\intx   Z^{i}g_{i}
$
plays the role of a generic Hamiltonian and where the generating functions
$g_{i}$ embody the arbitrariness in the variations. In particular, for the MHD
model, in terms of the density variables introduced in Ref.~\cite{amp1},  
where, as defined in Sec.~\ref{sec:lag},  $\mathbf{M}=\rho\mathbf{v}$,  the momentum density,  
and $\sigma=\rho
s$,  the entropy per unit volume, the functional that generates the
dynamically accessible variations can be written as%
\begin{equation}
G=\intx \left(  \mathbf{g}_{1}\cdot\mathbf{M}+g_{2}\sigma+g_{3}%
\rho+\mathbf{g}_{4}\cdot\mathbf{B}\right)
\label{eq:gen_fun}%
\end{equation}
and the Poisson bracket is of Lie-Poisson form \cite{morrison-greene,morrison-greene-cor}
\begin{align}
\left\{  F,G\right\}   &  =-%
\intx
\left[  \rho\left(  F_{\mathbf{M}}\cdot\mathbf{\nabla}G_{\rho}-G_{\mathbf{M}%
}\cdot\mathbf{\nabla}F_{\rho}\right)  \right. \nonumber\\
&  +\mathbf{M}\cdot\left[  \left(  F_{\mathbf{M}}\cdot\mathbf{\nabla}\right)
G_{\mathbf{M}}-\left(  G_{\mathbf{M}}\cdot\mathbf{\nabla}\right)
F_{\mathbf{M}}\right] \nonumber\\
&  +\sigma\left(  F_{\mathbf{M}}\cdot\mathbf{\nabla}G_{\sigma}-G_{\mathbf{M}%
}\cdot\mathbf{\nabla}F_{\sigma}\right) \nonumber\\
&  \left.  +F_{\mathbf{M}}\cdot\left(  \mathbf{B}\times\left(  \mathbf{\nabla
}\times G_{\mathbf{B}}\right)  \right)  +F_{\mathbf{B}}\cdot\mathbf{\nabla
}\times\left(  \mathbf{B}\times G_{\mathbf{M}}\right)  \right]  ,
\nonumber
\end{align}
i.e., linear with respect to each variable. (Note, for compactness we have written the  bracket of  \cite{morrison-greene};  for equilibria with $\nabla\cdot \bfB_e=0$ results are identical to those with the more general bracket of \cite{morrison-greene-cor}.)

Stability is thus given by the positiveness of the second dynamically
accessible variation of the Hamiltonian%
\bq
H=\intx\left(  \frac{M^{2}}{2\rho}+\rho U+\frac{B^{2}}{8\pi}\right).
\eq
The first order dynamically accessible variations result
\begin{align}
\delta\rho_{\mathrm{da}}  &  =\mathbf{\nabla}\cdot\left(  \rho\mathbf{g}_{1}\right)  ,
\label{eq:var_0_1}\\
\delta\mathbf{M}_{\mathrm{da}}  &  =\rho\mathbf{\nabla}g_{3} +\left(
\mathbf{\nabla}\times\mathbf{M}\right)  \times\mathbf{g}_{1}+\mathbf{M\nabla
}\cdot\mathbf{g}_{1}\nonumber\\
&  +\mathbf{\nabla}\left(  \mathbf{M}\cdot\mathbf{g}_{1}\right)
+\sigma\mathbf{\nabla}g_{2}+\mathbf{B}\times\left(  \mathbf{\nabla}%
\times\mathbf{g}_{4}\right)  ,\label{eq:var_0_2}\\
\delta\sigma_{\mathrm{da}}  &  =\mathbf{\nabla}\cdot\left(  \sigma
\mathbf{g}_{1}\right)  ,\label{eq:var_0_3}\\
\delta\mathbf{B}_{\mathrm{da}}  &  =\mathbf{\nabla}\times\left(
\mathbf{B}\times\mathbf{g}_{1}\right)  
\label{eq:var_0_4}%
\end{align}
and the first variation of the Hamiltonian can be written as%
\bqy
\delta H_{\mathrm{da}}&=&\intx
\bigg[  \frac{\mathbf{M}}{\rho}\cdot\delta\mathbf{M}_{\mathrm{da}}
+T\delta
\sigma_{\mathrm{da}}+\frac{1}{4\pi}\mathbf{B}\cdot\delta\mathbf{B}%
_{\mathrm{da}}
\nonumber\\
&{\ }&
+ \left(  -\frac{M^{2}}{2\rho^{2}}+U+\frac
{p}{\rho}-\frac{\sigma}{\rho}T\right)  \delta\rho_{\mathrm{da}}
\bigg]  . \label{eq:d1H}%
\eqy
By inserting into Eq. (\ref{eq:d1H}) the expressions obtained for the
dynamically accessible variations, Eqs. (\ref{eq:var_0_1})-(\ref{eq:var_0_4}),
we get the set of MHD equilibrium equations.

The second variation of the Hamiltonian results%
\begin{align}
\delta^{2}H_{\mathrm{da}}  &  =\intx
\bigg[
\frac{1}{\rho}
\left|
\delta\mathbf{M}_{\mathrm{da}}
\right|^{2}
- 2\frac{\mathbf{M}}{\rho^{2}}
\cdot\delta\mathbf{M}_{\mathrm{da}}\delta\rho_{\mathrm{da}}
\nonumber\\
&\hspace{-.9 cm} +
\left(
\frac{M^{2}}{\rho^{3}} +\rho \frac{\p^2 U}{\p \rho^2}
+ 2\frac{\p U}{\p \rho} - 2\frac{\sigma}{\rho}\frac{\p^2 U}{\p \rho \p s}
+\frac{\sigma^{2}}{\rho^{3}} \frac{\p^2 U}{\p s^2}
\right)
\left(\delta\rho_{\mathrm{da}}\right)^{2}
\nonumber\\
&  \hspace{-.2cm}+ 2\left(
\frac{\p^2 U}{\p \rho \p s} -\frac{\sigma}{\rho^{2}}  \frac{\p^2 U}{\p s^2}  
\right)
\delta \rho_{\mathrm{da}}\delta\sigma_{\mathrm{da}}+\frac{1}{\rho}  \frac{\p^2 U}{\p s^2}
\left(
\delta\sigma_{\mathrm{da}}
\right)^{2}
\nonumber\\
& \hspace{-.3cm}
+ \frac{1}{4\pi}
\left|\delta
\mathbf{B}_{\mathrm{da}}\right|^{2}
 +   \frac{2\mathbf{M}}{\rho}\cdot\delta^{2} \mathbf{M}_{\mathrm{da}}
 +2\frac{\p U}{\p s}\delta^{2}\sigma_{\mathrm{da}}
 \label{eq:Hvar_sec}
 \\
&\hspace{-.5 cm}
 +2\left(\rho \frac{\p U}{\p \rho}+U-\frac{\sigma}{\rho}
\frac{\p U}{\p s} -\frac{M^{2}}{2\rho^{2}}\right)  \delta^{2}\rho_{\mathrm{da}}
 + \frac{\mathbf{B}}{2\pi}\cdot\delta^{2}\mathbf{B}_{\mathrm{da}}
\bigg]
\nonumber
\end{align}
where $\delta^{2}Z_{\mathrm{da}}$ are the second order variations obtained as%
\begin{equation}
\delta^{2}Z_{\mathrm{da}}=\left\{  G^{\left(  2\right)  },Z\right\}  +\frac
{1}{2}\left\{  G^{\left(  1\right)  },\left\{  G^{\left(  1\right)
},Z\right\}  \right\}
\end{equation}
Notice that $G^{\left(  1\right)  }$ represents the first order generating
functional, i.e. the functional (\ref{eq:gen_fun}), whereas $G^{\left(
2\right)  }$ is a second order functional. However, it is easy to show that
altogether the terms corresponding to this second functional become null at
the equilibrium points.

The second order variations result%
\begin{align}
\delta^{2}\rho_{\mathrm{da}}  & =\frac{1}{2}\mathbf{\nabla}\cdot\left(
\delta\rho_{\mathrm{da}}\mathbf{g}_{1}\right)
\nonumber
\\
\delta^{2}\mathbf{M}_{\mathrm{da}}  &  =\frac{1}{2}\left[  \delta
\rho_{\mathrm{da}}\mathbf{\nabla}g_{3}+\left(  \mathbf{\nabla}\times
\delta\mathbf{M}_{\mathrm{da}}\right)  \times\mathbf{g}_{1}+\delta
\mathbf{M}_{\mathrm{da}}\mathbf{\nabla}\cdot\mathbf{g}_{1}\right. \nonumber\\
&  +\left.  \mathbf{\nabla}\left(  \delta\mathbf{M}_{\mathrm{da}}%
\cdot\mathbf{g}_{1}\right)  +\delta\sigma_{\mathrm{da}}\mathbf{\nabla}%
g_{2}+\delta\mathbf{B}_{\mathrm{da}}\times\left(  \mathbf{\nabla}%
\times\mathbf{g}_{4}\right)  \right]
\nonumber
\\
\delta^{2}\sigma_{\mathrm{da}}  &  =\frac{1}{2}\mathbf{\nabla}\cdot\left(
\delta\sigma_{\mathrm{da}}\mathbf{g}_{1}\right)
\nonumber
\\
\delta^{2}\mathbf{B}_{\mathrm{da}}  &  =\frac{1}{2}\mathbf{\nabla}%
\times\left(  \delta\mathbf{B}_{\mathrm{da}}\times\mathbf{g}_{1}\right)
\nonumber
\end{align}
In order to compare the results of dynamical accessible variations with those
of the Lagrangian approach (see Sec. \ref{sec:lag}), we introduce a
\textquotedblleft Lagrangian" velocity variation defined as
$
\delta\mathbf{v}_{\mathrm{la}}= \partial\boldsymbol{\eta}/{\partial t} 
-\boldsymbol{\eta}\cdot \nabla \mathbf{v} 
+\mathbf{v}\cdot\nabla\boldsymbol{\eta}
$.
After some manipulations, we obtain%
\bqy
\delta^{2}H_{\mathrm{da}}\left[  \mathbf{g}
\right]
&=&\intx
\rho\big|
 \delta\mathbf{v}_{\mathrm{da}} -\mathbf{g}_{1}\cdot\mathbf{\nabla v} 
 +\mathbf{v}\cdot \nabla\mathbf{g}_{1}\big| ^{2}
\nonumber\\
&{\ }& \qquad + \delta W_{\mathrm{la}}\left[ \mathbf{g}_{1}\right]  ,
\label{eq:d2H_da}%
\eqy
where $\mathbf{g}:=(\mathbf{g}_{1},g_{2},g_{3},\mathbf{g}_{4})$,
\bqy
\delta\mathbf{v}_{\mathrm{da}}&=&\mathbf{\nabla}\left(  g_{3} 
+\mathbf{g}_{1}\cdot\mathbf{v}\right)  - \mathbf{g}_{1}\times\left( \nabla
\times\mathbf{v}\right)
\nonumber\\
&{\ }& \qquad  + \frac{\sigma}{\rho} \nabla g_{2} 
+\frac{1}{\rho}\mathbf{B}\times\left(   {\nabla}\times\mathbf{g}_{4}\right)
\nonumber
\eqy
and the quadratic form $\delta W_{\mathrm{la}}$ is now expressed in terms of
$\mathbf{g}_{1}$. Notice that the variation $\delta^{2}H_{\mathrm{da}}$ is
formally identical to the variation $\delta^{2}H$ obtained in the Lagrangian
description, where $\eta=-\mathbf{g}_{1}$ and $\delta\mathbf{v}_{\mathrm{da}}$
replaces $\delta\mathbf{v}_{\mathrm{la}}$.  As shown in Sec. \ref{ssec:equil-stab}  in the Lagrangian description the integrand in the first term on the r.h.s. of Eq. (\ref{eq:d2H_da})  can be re-expressed in terms of 
 $\left(  \boldsymbol{\pi}_{\eta}%
-\rho_e\mathbf{v}_e \cdot \nabla \boldsymbol{\eta}\right) = {\partial\boldsymbol{\eta}}/{\partial t}$ and an arbitrary variation $ \boldsymbol{\pi}_{\eta}$ can make this term null.  On the contrary, in the case of  dynamically accessible  perturbations  the arbitrariness of the
variation $\delta\mathbf{v}_{\mathrm{da}}$ is described by the functions
$\mathbf{g}_{1}$, $g_{2}$, $g_{3}$ and $\mathbf{g}_{4}$ and the first term of
$\delta^{2}H_{\mathrm{da}}$ can be written as
\bqy
{\Delta}&=&\intx \rho
\big|  \delta\mathbf{v}_{\mathrm{da}}
- \mathbf{g}_{1} \cdot \nabla \mathbf{v} 
+ \mathbf{v}\cdot \nabla \mathbf{g}_{1} \big|^{2}
\nonumber\\
&=&
\intx \rho\Big|  {\nabla}g_{3}
+ \mathbf{v}\times\left(
\nabla\times\mathbf{g}_{1}\right)
+ 2\left(  \mathbf{v}\cdot\nabla\right) \mathbf{g}_{1}
\nonumber\\
&{\ }&
\hspace{1.5 cm} +\frac{\sigma}{\rho} {\nabla}g_{2}
+ \frac{1}{\rho} \mathbf{B}\times\left( {\nabla}\times\mathbf{g}_{4}\right)
\Big|^{2} .
\label{eq:term}%
\eqy
The functions $g_{2}$, $g_{3}$ and $\mathbf{g}_{4}$ inside $\Delta$,  which is  non-negative,  do not  appear inside the last term in  Eq. (\ref{eq:d2H_da}). Thus one  might  think that a suitable choice of these functions exists that  yields $\Delta=0$. However  such a choice can be made only if specific solvability conditions (analogous to those of the magnetic differential equations discussed by Newcomb in Ref.~\cite{newcomb59}) are satisfied. Thus, in general, we can only try to minimize $\Delta$ with respect to the functions $g_{2}$, $g_{3}$ and $\mathbf{g}_{4}$.  This kind of minimization was first suggested  in the dynamically accessible context applied to Vlasov theory in Ref.~\cite{Morrison1990}, but  a  similar procedure was adopted for MHD   equilibrium  configurations with  nested flux surfaces  in  \cite{hameiri3}  and  without flux surfaces  in   \cite{hameiri2}.   In the following we are going to analyze the symmetric case, which represents a
good  benchmark for the procedure and permits a direct comparison with the results obtained in Sec. \ref{sec:ec}.

The first variation of ${\Delta}$\ with respect to $g_{2,3}$,  and
$\mathbf{g}_{4}$  yields
$
{\delta{\Delta}}/{\delta g_{2}}   =\mathbf{\nabla}
\cdot\left(\sigma\mathbf{X}\right)
$,
$
{\delta{\Delta}}/{\delta g_{3}}=\mathbf{\nabla}
\cdot\left( \rho\mathbf{X}\right)
$, and
$ {\delta{\Delta}}/{\delta\mathbf{g}_{4}}
=\mathbf{\nabla}\times
\left(\mathbf{X}\times\mathbf{B}\right)
$,
where we defined the vector field $\mathbf{X}$\ as
\bqy
\mathbf{X}&:=& {\nabla}g_{3}+\mathbf{v}\times\left(
\mathbf{\nabla}\times\mathbf{g}_{1}\right)  
+ 2 \left(  \mathbf{v\cdot\nabla }\right)  \mathbf{g}_{1}
\nonumber\\
&{\ }&
+\frac{\sigma}{\rho} {\nabla}g_{2} 
+ \frac{1}{\rho}\mathbf{B}\times\left( {\nabla}\times\mathbf{g}_{4}\right)  .
\label{eq:X}%
\eqy
The minimum of term (\ref{eq:term}) satisfies%
\begin{align}
{\nabla}\cdot\left(  \rho\mathbf{X}\right)   &  =0 \quad 
\Leftrightarrow\quad  {\nabla}\cdot\left( \de \mathbf{M}_{\mathrm{da}}\right)   =0,
\label{eq:x1}\\
\mathbf{X}\cdot {\nabla}\frac{\sigma}{\rho}  &  =0 
\quad 
\Leftrightarrow\quad  {\nabla}\cdot\left( \de (\mathbf{M}\,\sigma/\rho)_{\mathrm{da}}\right)   =0,
\label{eq:x2}\\
{\nabla}\times\left(  \mathbf{X}\times\mathbf{B}\right)   &  =0
\quad 
\Leftrightarrow\quad  {\nabla}\times\left( \de \mathbf{E}_{\mathrm{da}}\right)   =0.
\label{eq:x3}%
\end{align}
where $\mathbf{E}= -\mathbf{v}\times \mathbf{B}/c$ and the equivalencies above, which are new and give insight,  can be ascertained by a straightforward calculation.   By considering symmetric configurations, where
$\mathbf{B}=B_{h}\mathbf{h} + {\nabla}\psi\times k\mathbf{h}$ and
$\frac{\sigma}{\rho}=s\left(  \psi\right)$,
we obtain $\mathbf{X}=X_{h}\mathbf{h}+\frac{1}{\rho}\mathbf{\nabla}\chi\times
k\mathbf{h}$ from Eq.~(\ref{eq:x1}),
$
 {\nabla}\chi\times {\nabla}\psi\cdot k\mathbf{h}=0
\rightarrow \chi\left(  \psi\right)
$
from Eq.~(\ref{eq:x2}),  and
\[
{\nabla}\times\left[  k\left(  X_{h}-\frac{\chi^{\prime}}{\rho} B_{h}\right)
 {\nabla}\psi\right] = 0\quad\rightarrow
\]
$X_{h}=\frac{\chi^{\prime}}{\rho}B_{h} +\frac{1}{k}G\left(  \psi\right)$ from Eq.~(\ref{eq:x3}).

Thus, for symmetric configurations, the vector field $\mathbf{X}$  that
minimizes the term $\Delta$  can be written as%
\begin{equation}
\mathbf{X}_{\min}=\frac{F}{\rho}\mathbf{B}+G\frac{\mathbf{h}}{k},
\label{eq:Xmin}%
\end{equation}
where $F=\chi^{\prime}$\ and $G$\ are two generic functions of $\psi$.

Then, we consider the symmetric version of Eq.~(\ref{eq:X}), which yields%
\begin{align}
\mathbf{X}  & :=  {\nabla}\left(  g_{3} + \frac{\sigma}{\rho}g_{2}\right)
 -\left(  k\mathbf{h}\cdot\frac{1}{\rho} {\nabla} \times\mathbf{g}_{4} 
 + s^{\prime}g_{2}\right)   {\nabla}\psi
\nonumber\\
& + \frac{kB_{h}}{\rho} {\nabla}\frac{{g}_{4h}}{k}-k\mathbf{h}\left(  \frac{1}{\rho
}\mathbf{B}_{\bot}\cdot\mathbf{\nabla}\frac{g_{4h}}{k}\right)
 +\frac{v_{h}}{k}\mathbf{\nabla}\left(kg_{1h}\right)
\nonumber\\
&  +\frac{v_{h}}%
{k} {\nabla}\times\left(  k\mathbf{h}\right)  \times\mathbf{g}_{1\bot} 
- 2v_{h}\mathbf{h}\left(  \mathbf{g}_{1}\cdot\frac{1}{k} {\nabla}k\right)
\nonumber\\
&  +\mathbf{v}_{\bot}\times\left( {\nabla}\times\mathbf{g}_{1}\right) 
+ 2\left(  \mathbf{v}_{\bot}\cdot {\nabla}\right)
\mathbf{g}_{1}
 \label{eq:Xsym}%
\end{align}
and, by combining  Eq.~(\ref{eq:Xmin}) and Eq.~(\ref{eq:Xsym}), we obtain
\begin{align}
\frac{F}{\rho}B_{h} + G\frac{1}{k}&=-k\mathbf{B}_{\bot}
\cdot\frac{1}{\rho} {\nabla}\frac{g_{4h}}{k} 
- 2\frac{v_{h}}{k}\mathbf{g}_{1}\cdot {\nabla}k
\nonumber\\
& +\mathbf{v}_{\bot}\cdot\left[  \frac{1}{k} {\nabla}\left(  kg_{1h}\right)  
+ \frac{1}{k} {\nabla}\times\left( k\mathbf{h}\right)  \times\mathbf{g}_{1\bot}\right]
\nonumber
\end{align}
and%
\begin{align}
\frac{F}{\rho} {\nabla}\psi\times k\mathbf{h}  &  = 
{\nabla}\left(  g_{3}+\frac{\sigma}{\rho}g_{2}\right)   
+ \frac{kB_{h}}{\rho} {\nabla}\frac{g_{4h}}{k}
 + \frac{v_{h}}{k} {\nabla}\left(  kg_{1h}\right)
\nonumber\\
& \hspace{-1 cm}
-\left(  k\mathbf{h}\cdot
\frac{1}{\rho} {\nabla}\times\mathbf{g}_{4} 
+ s^{\prime}g_{2}\right)
 {\nabla}\psi
 + \frac{v_{h}}{k} {\nabla}\times\left( k\mathbf{h}\right)  \times\mathbf{g}_{1\bot}
\nonumber\\
& + \mathbf{v}_{\bot}\times\left(   {\nabla}\times\mathbf{g}_{1\bot}\right)  
+ 2 \left(  \mathbf{v}_{\bot}\cdot {\nabla}\right)\mathbf{g}_{1\bot}\,.
\nonumber
\end{align}

Now, in order to determine $F$\ and $G$, we multiply Eq.~(\ref{eq:Xsym}) by
$\rho\mathbf{h}/k$\ and we integrate this expression in a domain\ $\Psi
$\ bounded by two magnetic flux surfaces $\psi$\ and $\psi+d\psi$
\begin{equation}
\int_{\Psi}\!d^{3}x\,  F\frac{B_{h}}{k}  +\int_{\psi}\!d^3x \,
\frac{G}{k^{2}} =\int_{\Psi}\!d^3x\,   \rho\frac{X_{h}}{k} \,,
\nonumber
\end{equation}
and we obtain
\bqy
F\left\langle \frac{B_{h}}{k}\right\rangle &+& G\left\langle \frac{1}{k^{2}%
}\right\rangle =\Big\langle -2\frac{\rho v_{h}}{k^{2}}\mathbf{g}_{1}%
\cdot\mathbf{\nabla}k
\label{eq:sys1}\\
&{\ }&  +\frac{\rho\mathbf{v}_{\bot}}{k^{2}}\cdot\left[
 {\nabla}\left(  kg_{1h}\right)  + {\nabla}\times\left(
k\mathbf{h}\right)  \times\mathbf{g}_{1\bot}\right]  \Big\rangle
\nonumber%
\eqy
where $\left\langle f\right\rangle =\int_{\psi}\frac{d^{2}x}{\left\vert
 {\nabla}\psi\right\vert}\, f$
indicates the surface integral on a flux surface. In order to obtain Eq.~(\ref{eq:sys1})
we used the fact that $F=F\left(  \psi\right)$ and
$G=G\left(  \psi\right)$ and the equation%
\begin{equation}
\int_{\Psi}\!d^3x\, \left[  \mathbf{B}_{\bot}\cdot {\nabla}\frac{g_{4h}}%
{k}\right]  =\int_{\Psi}\!d^3x\,  {\nabla}\cdot\left[  \mathbf{B}_{\bot}\frac{g_{4h}}{k}\right]   =0,
\nonumber
\end{equation}
where the last equality follows from the fact that the boundaries are flux surfaces.

Next, we multiply Eq.~(\ref{eq:Xsym}) by $\mathbf{B}$\ and again we integrate
in $\Psi$ to obtain
\begin{equation}
F\left\langle \frac{\left\vert \mathbf{B}\right\vert ^{2}}{\rho}\right\rangle
+ G \left\langle \frac{B_{h}}{k}\right\rangle =\big\langle 2\mathbf{B}%
\cdot\left(  \mathbf{v}\cdot\nabla \mathbf{g}_{1}\right)  \big\rangle ,
\label{eq:sys2}%
\end{equation}
where we use the expressions $\mathbf{B}\cdot {\nabla}\psi=0$,
\begin{equation}
\int_{\Psi}\!d^3x\, \big[  \mathbf{B}\cdot\mathbf{v}\times\left(  \mathbf{\nabla} \times\mathbf{g}_{1}\right)  \big]  
= \int_{\Psi}\!d^3x\, \mathbf{\nabla}\cdot 
\big[ 
 \mathbf{g}_{1}\times\left(  \mathbf{B}\times\mathbf{v}\right)
\big] \, ,
\nonumber
\end{equation}
and $\int_{\Psi}\!d^3x\,  \mathbf{B}\cdot  {\nabla} g=\int_{\Psi}\!d^3x\,   {\nabla}\cdot(\mathbf{B} g)$ for
$g= g_{3} +{\sigma g_{2}}/{\rho}$.

The two Eqs.~(\ref{eq:sys1}) and (\ref{eq:sys2}) can be rewritten as  $ \mathbb{A}\cdot \boldsymbol{\Xi}  = \boldsymbol{\Gamma}$, 
where $\boldsymbol{\Xi} =\left(  F,G\right)^{t}$,  the $2\times 2$ matrix $ \mathbb{A}$  is
\begin{equation}
 \mathbb{A}=%
\begin{bmatrix}
\left\langle \frac{\left\vert \mathbf{B}\right\vert ^{2}}{\rho}\right\rangle
& \left\langle \frac{B_{h}}{k}\right\rangle \\
\left\langle \frac{B_{h}}{k}\right\rangle  & \left\langle \frac{1}{k^{2}%
}\right\rangle
\end{bmatrix}
\end{equation}
and the vector $\boldsymbol{\Gamma}$ is %
\begin{equation}
\boldsymbol{\Gamma} =%
\begin{bmatrix}
\left\langle 2\mathbf{B}\cdot\left(  \mathbf{v}\cdot\mathbf{\nabla g}%
_{1}\right)  \right\rangle \\
\left\langle
\frac{\rho}{k^{2}} \, \mathbf{v}_{\bot}\cdot \mathcal{L}_k\, \mathbf{g}_{1}
 -\frac{2\rho}{k^{2}} \, v_{h} \mathbf{g}_{1}\cdot {\nabla} k
  \right\rangle
\end{bmatrix}
^{t}%
\end{equation}
where $
\mathcal{L}_k\,\mathbf{g}_1:=   {\nabla}\left(
kg_{1h}\right)  + {\nabla}\times\left(  k\mathbf{h}\right)
\times\mathbf{g}_{1\bot}\,.
$
Moreover, we notice that the coefficients of $ \mathbb{A}$\ depend only on the
equilibrium fields, while the vector $\boldsymbol{\Gamma} $\ depends also on $\mathbf{g}_{1}$.
Finally, we  solve the linear system $ \mathbb{A}\cdot \boldsymbol{\Xi} =\boldsymbol{\Gamma}$ as $\boldsymbol{\Xi} = \mathbb{A}^{-1}\cdot\boldsymbol{\Gamma}$, 
and then substitute  the solution
$
\mathbf{X}_{\mathrm{min}}=\Xi_{1}{\mathbf{B}}/{\rho}+\Xi_{2}{\mathbf{h}}/{k},
$
into Eq.~(\ref{eq:term}) to obtain%
\bqy
\Delta_{\mathrm{min}}&=&\intx \rho\left\vert \mathbf{X}_{\mathrm{min}}\right\vert ^{2}
\nonumber\\
&=&\intx
\left(
\Xi_{1}^{2}\, \frac{\left\vert \mathbf{B}\right\vert ^{2}}{\rho}
+2\, \Xi_{1}\Xi_{2} \, \frac{B_{h}}{k} +\Xi_{2}^{2}\, \frac{1}{k^{2}}\right) 
\nonumber\\
&=&\int_{\psi} d\psi\,\,     \boldsymbol{\Xi }^t\cdot \mathbb{A}\cdot \boldsymbol{\Xi }
=\int_{\psi}d\psi \,\,    \boldsymbol{\Gamma}^t \cdot  \mathbb{A}^{-1}\!\cdot \boldsymbol{\Gamma}  \,. 
\label{eq:term_min}
\eqy
It can be shown that the solvability conditions of Eq. (\ref{eq:term}) correspond to $\boldsymbol{\Gamma} =0$.  
In this case the condition $ \delta^{2}H_{\mathrm{da}}>0$ corresponds to $\delta^{2}W_{\mathrm{la}}>0$, i.e., dynamically accessible  stability conditions are equivalent to those obtained in Ref.~\cite{Frieman1960}. 

\bigskip


\section{Summary and conclusions}
\label{sec:conclu}

In this paper we have described various forms of stability for MHD within the Hamiltonian framework, which is an  efficacious  stability framework because the Hamiltonian can serve as a Lyapunov functional.  

We first described the Hamiltonian structure in terms of the Lagrangian variables, which being particle-like naturally has the canonical Hamiltonian form.  We then described how time-dependent relabeling is a canonical transformation that amounts to a local frame change that can be used to remove the time dependence of fluid element trajectories that occurs on the Lagrangian variable level for stationary (time-independent) Eulerian equilibria.   For MHD in \cite{Frieman1960} and also in other works in the kinetic theory context (e.g.~\cite{morrison-pfirsch}),  time dependence was removed by measuring the displacement relative to  the equilibrium trajectory.   This can be viewed as a linear ramification of our fully nonlinear relabeling development, which to our knowledge is new.   We also discussed the Hamiltonian in the relabeled frame and compared it to that for global transformations such as occurs for the frame shift corresponding to the total momentum. Then, the interrelationship  between relabeling and the Euler-Lagrange map was described for equilibrium states. With these tools at hand we were able to arrive at an energy expression that was compared to that of  \cite{Frieman1960}.

Next we described Hamiltonian stability on the Eulerian variable level.  This was done within the confines of a formulation that represents  general symmetry, which affords a rich Casimir structure for ascertaining stability within various symmetry classes. General sufficient conditions for stability were obtained by incisive analysis of the energy-Casimir functional. 

Finally, the dynamically accessible variations, based on the theory introduced in \cite{morrison-pfirsch,Morrison1990} and developed in generality in \cite{Kandrup1993,{morrison98}}, were employed.  This allowed the investigation of arbitrary equilibria without the imposition of symmetry.   Extremization of the energy functional was done as in \cite{hameiri2,morrison-pfirsch,Morrison1990} and stability under this kind of constraint was determined. 

 As pointed out in \cite{morrison98},  differences in the various stability conditions arise  because different representations of a theory can incorporate different constraints.  In closing we make a few comments on the comparisons between the various stability results,  leaving  more in-depth comparisons to the companion paper \cite{amp2b},  where specific examples will be treated in detail. 
 
First consider the development  of Sec.~\ref{sec:lag} in terms of Lagrangian variables. Although our sufficient conditions for stability are  the same as those  of  \cite{Frieman1960}, the manner of derivation and meaning are different.  In  \cite{Frieman1960} the stability conditions are obtained by manipulation of the linear equations of motion and subsequent analysis based on the  insertion of exponential time dependence.  However, our development is purely Hamiltonian:  it  proceeds by expansion of the  fully nonlinear invariant energy, in the manner of Lagrange and Dirichlet of usual Hamilton theory, and no assumption is made about the temporal behavior of the solution.    It is important to realize  that linear equations of motion can have more than one quadratic invariant, and such invariants need not be the expansion of  an invariant of the  nonlinear system.  For finite-dimensional  systems,  definiteness of the expansion of the Hamiltonian to second order actually implies nonlinear stability, i.e., stability  under the full nonlinear dynamics. However,  stability based on the definiteness of an invariant obtained by manipulation of a  linear equation of motion is significantly weaker.  In fact, it is possible that systems shown to be stable by such a procedure can in fact be unstable to arbitrarily small perturbations.  (See Sec.~VI of \cite{morrison98} for discussion.)  For infinite-dimensional systems,  definiteness of $\de^2H$ of (\ref{eq:d2H_Lagr}) is a step toward a proof of nonlinear stability.   However,   rigorous proofs of stability can be quite subtle and difficult; since  stability is  norm dependent,   functional analysis is unavoidable (see, e.g., Refs.~\cite{Batt, Rein}). 

Direct  comparison of the stability conditions of Secs.~\ref{sec:lag}, \ref{sec:ec}, and \ref{sec:da} is complicated by the fact that not all apply to the same equilibria.  Although the Lagrangian and dynamically accessible methods apply to general equilibria, the energy-Casimir results as developed only apply to symmetric equilibria. Consequently, our comparisons below will implicitly assume equivalent equilibria. 

Let us denoted by $\mathfrak{P}:=\{\delta \rho, \delta \mathbf{M}, \delta \si, \delta\mathbf{B}\}$ the set of first order unconstrained perturbations of the Eulerian variables,  i.e.,  the perturbed variables $\delta \rho$, $\delta \mathbf{v}$, etc.\   are arbitrary and completely independent of each other.  This is the largest set of perturbations.  The set $\mathfrak{P}_{\mathrm{ec}}$ used in Sec.~\ref{sec:ec} is similarly unconstrained,  except within our  symmetry class we have built in $\nabla\cdot\delta B_{\mathrm{ec}}=0$. The set $\mathfrak{P}_{\mathrm{ec}}$ is the largest of this paper.  Upon comparing  Eqs.~(\ref{eq:laDrho})--(\ref{eq:laDB}) with Eqs.~(\ref{eq:var_0_1})--(\ref{eq:var_0_4}), 
we see that with the identification  $\mathbf{g}_1\equiv-\boldsymbol{\eta}$,  the sets $\mathfrak{P}_{\mathrm{la}}$ and  $\mathfrak{P}_{\mathrm{da}}$ have all equivalent elements except for the momentum  perturbations, where $\delta \mathbf{M}_{\mathrm{la}}$ is given by (\ref{eq:laDM})   and $\delta \mathbf{M}_{\mathrm{da}}$ is given by (\ref{eq:var_0_2}). It is easy to see that dynamically accessible variations are less general than Lagrangian variations.  Because of the freedom to choose $\boldsymbol{\pi}_{\eta}$ in (\ref{eq:laDM}) at will, $\delta \mathbf{M}_{\mathrm{la}}$ is completely arbitrary.  However, to see that this is not the case for $\delta \mathbf{M}_{\mathrm{da}}$, consider the special case of static, uniform,  hydrodynamic equilibria,  where $\rho_e$= constant, $\mathbf{M}_e\equiv 0$, $\si_e$= constant, and $\mathbf{B}_e\equiv 0$, in which case 
\begin{equation}
 \delta \mathbf{v}_{\mathrm{da}}= \nabla g_3 + \frac{\si_e}{\rho_e} \nabla g_2= \nabla( g_3 +  \frac{\si_e}{\rho_e}   g_2)\,. 
 \end{equation}
 Thus, because  $\nabla \times \delta \mathbf{v}_{\mathrm{da}}\equiv 0$, this kind of perturbation is incapable of introducing vorticity into such a static fluid, in contrast to $ \delta \mathbf{v}_{\mathrm{la}}$.  For more general equilibria the    constraints implied by dynamical accessibility are more subtle and these will be considered on a case by case basis in \cite{amp2b}.  However,  in general the following is true:  
\[
\mathfrak{P}_{\mathrm{da}}\subset\mathfrak{P}_{\mathrm{la}}
\subset\mathfrak{P}_{\mathrm{ec}}\,. 
\]
As a side note, we observe that the expression $\delta \mathbf{v}_{\mathrm{da}}$, with $ \mathbf{v}\equiv 0$ and $\si$= constant,  is identical to the Clebsch representation introduced  in \cite{morrison82}.  Thus, this Clebsch representation is not capable of expressing all vector fields.

Given that dynamically accessible perturbations are constrained,  one must make a  decision based on the physics of the situation  to determine which  kinds of perturbations are relevant, an idea  that  was emphasized in \cite{morrison-pfirsch},  where the notion of  dynamical accessible stability was introduced,  and also  in subsequent work.   For example, if one is interested in ideal perturbations of a normal, fluid, i.e., the case where  viscosity is not important,  and it is assumed that the walls containing the fluid do not move normal to themselves, i.e., there is no stirring mechanism,   then  there is no physical mechanism by which vorticity can be introduced into the fluid, and we have a situation consistent with the case described above.  However,  if nondynamically accessible perturbations are important, then one might want to reassess the completeness  of the dynamical system governing the  phenomena.   If  nondynamically accessible  perturbations are allowed,  then one  might want a dynamical system that  reflects  their evolution in time. 
 
 It is not enough to just consider the first order perturbations: one must consider the energy expressions to which they correspond.   The perturbations  $\mathfrak{P}_{\mathrm{la}}$ are  to be  introduced into   (\ref{eq:d2H_Lagr}),   while the perturbations  $\mathfrak{P}_{\mathrm{da}}$ go into  (\ref{eq:d2H_da}).  Since $\boldsymbol{\pi}_{\eta}$, is arbitrary  the first term of (\ref{eq:d2H_Lagr}) was  made to vanish in Sec.~\ref{sec:lag},  leaving only $\delta^2 W_{\mathrm{la}}$.   while this was not the case when we analyzed  (\ref{eq:d2H_da})  in Sec.~\ref{sec:da}.   If one replaces $\boldsymbol{\pi}_{\eta}$ in (\ref{eq:d2H_Lagr}) by its dynamically accessible counterpart, 
 \[
 \delta \boldsymbol{\pi}_{\mathrm{da}} = - \nabla \mathbf{g}_1 \cdot \mathbf{M}_e - \si_e \nabla g_2 
- \rho_e \nabla g_3\,, 
  \]
 (see Eq.~(462) of \cite{morrison98}),  then one obtains  (\ref{eq:d2H_da}).  Thus,  the same energy expression applies to both, but in the dynamically accessible case one is constrained away from the minimum available in the Lagrangian case. Therefore,  to the extent that these expressions determine stability, Lagrangian stability implies dynamically accessible stability. 
 
 A comparison between the energy expressions used for  Lagrangian and energy-Casimir stability is also possible, if the former is restricted to symmetrical perturbations.    If one inserts for $\delta\mathcal{Z}_S$ in $\delta^2\mathcal{F}$ the Lagrangian induced symmetric variations of  Eqs.~(\ref{eq:laDrho})--(\ref{eq:laDB}), adapted for symmetry, then $\delta^2\mathcal{F}$ becomes identical to $\delta^2 H_{\mathrm{la}}$.  This calculation was done for a reduced system (compressible reduced MHD of \cite{hazeltine3})  in \cite{MTT}, but the calculation here for general symmetry is more complicated.   To effect this calculation,  (\ref{symB}) and (\ref{eq:laDB}) are used to obtain $\delta \psi_{\mathrm{la}} =-\boldsymbol{\eta}\cdot \nabla \psi_e$.  Beginning with the first term of (\ref{eq:d2F_diag}), with $\delta\mathbf{S}$ given by the first of Eqs.~(\ref{eq:varR}), and the perturbations $\delta \mathbf{B}$, $\delta \psi$, etc.\  replaced by their Lagrangian induced versions, we obtain 
\[
\intx
a_{1}\left|\delta\mathbf{S}\right|^{2} = 
\intx \rho\big| \delta \mathbf{v} + \boldsymbol{\eta}\cdot\nabla  \mathbf{v}_e - 
 \mathbf{v}_e \cdot\nabla \boldsymbol{\eta}\big|^2\,.
\]
 This calculation requires the removal of the functions $\mathcal{F}$ and $\mathcal{G}$ in lieu of $ \mathbf{v_e}$ by making use of   (\ref{eq:nP}) and the use of metric identities such as $\mathbf{h}\cdot\nabla \mathbf{h}\propto  { \boldsymbol{r}}$.  
The identity of the remaining portions of these energies follows similarly.  Given that Lagrangian perturbations are a subset of energy-Casimir variations, we conclude that energy-Casimir stability implies Lagrangian stability, as implied by these energy expressions.  Similarly, it was shown  in Sec.~VIC.2 of  \cite{morrison98} that  insertion of the  perturbations $\mathfrak{P}_{\mathrm{da}}$  into (\ref{eq:d2H_Lagr}) produces   (\ref{eq:d2H_da}), i.e.,  
$\delta^2\mathfrak{F}_\mathrm{da}\equiv\delta^2{H}_\mathrm{da}$ when the former is evaluated on first order dynamically accessible variations.  Thus,  we are led to  two conclusions:   
 \[
\mathfrak{stab}_{ec}\Rightarrow\mathfrak{stab}_{\mathrm{la}}\Rightarrow\mathfrak{stab}_{\mathrm{da}}\,,
\]
to the extent that each of these quadratic forms implies stability, and that all the quadratic forms are in fact the identical physical energy contained in a perturbation away from an equilibrium state, but how much of that energy can be tapped depends on the constraints embodied in the forms of the perturbations $\delta Z$. 
 
 \vspace{.2 cm}
 \noindent{\it Postcript:}  We wish to point out two references that were brought to our attention after the completion of this work. In  Ref.~\cite{yao}  the  authors  have used a form of Noether's theorem in an  action principle setting of MHD  to  compare Lagrangian and dynamically accessible perturbations, while in Ref.~\cite{moa} the author considers a case of energy-Casimir stability that is in the same vein as that of our Sec.~\ref{sec:ec}.


\section*{Acknowledgment}

\noindent PJM was supported by U.S. Dept.\ of Energy Contract \# DE-FG05-80ET-53088.


\end{document}